\documentclass{amsart}

\usepackage{amssymb,amsfonts,latexsym,amscd}
\usepackage{amsmath}
\def\gs{\Psi} %ground state of the full hamiltonian H
\def\pst{\gs^{\mathrm{trial}}} % trial function

\def\ce{\Phi}
\def\ceu{\Phi_1} %\cE_1 , \cE_2  and \cE_3  helps to define the trial state in paper I
\def\ced{\Phi_2}
\def\cet{\Phi_3}
\def\ceq{\Phi_4}
\def\cequ{\Phi_4^{(1)}}
\def\ceqd{\Phi_4^{(2)}}
\def\cedd{\tilde{\Phi}_2}
\def\ceddu{\tilde{\Phi}_2^{(1)}}
\def\ceddd{\tilde{\Phi}_2^{(2)}}
\def\ceddt{\tilde{\Phi}_2^{(3)}}
\def\cei{\ce_i}
\def\cej{\ce_j}
\def\vac{\Omega_f}  % vaccum in bosonic Fock space

\def\Aa{A^-} % annihilation part of the vector field operator
\def\Ac{A^+} % creation part of the vector field operator
\def\Htot{T} %Translation invariant Pauli-Fierz operator
\def\H{T(0)} %Transl. inv. Pauli-Fierz op. with total momentum 0

\newcommand{\gH}{{\mathfrak H}}  % Hilbert space
\newcommand{\gF}{{\mathfrak F}}  % Fock space

\def\C{{\mathbb{C}}} % Complex numbers
\def\N{{\mathbb{N}}} % Integers
\def\R{{\mathbb{R}}} % Real numbers
\newcommand{\pr}{{P_{\ced}}} % projection onto \ce_2

\def\Proj{\Gamma} % projection
\def\1{{\mathbf{1}}}
\def\la{\langle}
\def\ra{\rangle}
\renewcommand\d{\mathrm{d}}
\renewcommand\Re{\mathrm{Re}\,}
\renewcommand\Im{\mathrm{Im}\,}
\def\Or{\mathcal{O}}
\newtheorem{theorem}{Theorem}[section]

\newtheorem{corollary}{Corollary}[section]
\newtheorem{lemma}{Lemma}[section]

\newtheorem{proposition}{Proposition}[section]

\begin{document}

%%%%%%%%%%%%%%%%%%%%%%%%%%%%%%%%%%%%%%%%%%%%%%%%%%%%%%%%%%%%%
%%%%%%%%%%%%               TITLE         %%%%%%%%%%%%%%%%%%%%
%%%%%%%%%%%%%%%%%%%%%%%%%%%%%%%%%%%%%%%%%%%%%%%%%%%%%%%%%%%%%

\title[Non analyticity of the ground state energy in NRQED]
{Non analyticity of the ground state energy of the Hamiltonian for
Hydrogen atom in nonrelativistic QED}

\author{J.-M. Barbaroux$^1$, S.A. Vugalter$^2$,
}

\address{$^1$ Centre de Physique Th\'eorique, Luminy Case 907, 13288
 Marseille Cedex~9, France and D\'epartement de Math\'ematiques,
 Universit\'e du Sud Toulon-Var, 83957 La
 Garde Cedex, France}
\address{$^2$ Mathematisches Institut, Ludwig-Maximilians-Universit\"at
M\"unchen, Theresienstrasse 39, 80333 M\"unchen}
\email{barbarou@univ-tln.fr}
\email{wugalter@mathematik.uni-muenchen.de}

%%%%%%%%%%%%%%%%%%%%%%%   abstract   %%%%%%%%%%%%%%%%%%%%%%%%
\begin{abstract}
We derive the ground state energy up to the fourth order in the
fine structure constant $\alpha$ for the translation invariant
Pauli-Fierz Hamiltonian for a spinless electron coupled to the
quantized radiation field. As a consequence, we obtain the
non-analyticity of the ground state energy of the Pauli-Fierz
operator for a single particle in the Coulomb field of a nucleus.
\end{abstract}
%%%%%%%%%%%%%%%%%%%%   end of abstract    %%%%%%%%%%%%%%%%%%%

\maketitle

%%%%%%%%%%%%%%%%%%%%%%%%%%%%%%%%%%%%%%%%%%%%%%%%%%%%%%%%%%%%%
%%%%%%%               END OF TITLE     %%%%%%%%%%%%%%%%%%%%%%
%%%%%%%%%%%%%%%%%%%%%%%%%%%%%%%%%%%%%%%%%%%%%%%%%%%%%%%%%%%%%

%%%%%%%%%%%%%%%%%%%%%%%%%%%%%%%%%%%%%%%%%%%%%%%%%%%%%%%%%%%%%%%
%%%%%%%%%%%%%%%%%%%%%%%%%%%%%%%%%%%%%%%%%%%%%%%%%%%%%%%%%%%%%%%
%%%%%%%%%%     INTRODUCTION                %%%%%%%%%%%%%%%%%%%%
%%%%%%%%%%%%%%%%%%%%%%%%%%%%%%%%%%%%%%%%%%%%%%%%%%%%%%%%%%%%%%%
%%%%%%%%%%%%%%%%%%%%%%%%%%%%%%%%%%%%%%%%%%%%%%%%%%%%%%%%%%%%%%%

%%%%%%%%%%%%%%%%%%%%%%%%%%%%%%%%%%%%%%%%%%%%%%%%%%%%%%%%%%%%%%%
%%%%%%%%%%%%%%%%%%%%%%%%%%%%%%%%%%%%%%%%%%%%%%%%%%%%%%%%%%%%%%%
\section{Introduction}

We study the translation invariant Pauli-Fierz Hamiltonian
describing a spinless electron interacting with the quantized
electromagnetic radiation field.

In the last fifteen years, a large number of rigorous results were
obtained concerning the spectral properties of Pauli-Fierz
operators, starting with the pioneering works of Bach, Fr\"ohlich
and Sigal \cite{Bachetal1995, BFS1999, Bachetal1998}. In
particular, the ground state energy were intensively studied
(\cite{GLL}, \cite{LL}, \cite{HVV}, \cite{Chenetal2003},
\cite{BLV}, \cite{BCV0}, \cite{Hainzletal2005}, \cite{GrHa}).

One of the problems recently discussed is the existence of an
expansion in powers of the fine structure constant $\alpha$ for
the ground state energy of Pauli-Fierz operators. The very first
results in this direction are due to Pizzo \cite{Pizzo} and later
on Bach, Fr\" ohlich and Pizzo \cite{Bachetal2006}, where the
operator for the Hydrogen atom is considered. In
\cite{Bachetal2006}, a sophisticated rigorous renormalization
group analysis is developed in order to determine the ground state
energy, up to any arbitrary precision in powers of $\alpha$, with
an expansion of the form
 $$
  \varepsilon_0 + \sum_{k=1}^{2N} \varepsilon(k) \alpha^{k/2}\, +\,
  o(\alpha^N)\, ,
 $$
for any given $N$, where the coefficients $\varepsilon_k(\alpha)$
may diverge as $\alpha\to 0$, but are smaller in magnitude than
any power of $\alpha^{-1}$. The recursive algorithms developed in
\cite{Bachetal2006} are highly complex, and explicitly computing
the ground state energy to any subleading order of $\alpha$ is an
extensive task. In the physical model where the photon form factor
in the quantized electromagnetic vector potential contains the
critical frequency space singularity responsible for the infamous
infrared problem, it is expected that the rate of divergence of
some of these coefficient functions $\varepsilon_k(\alpha)$ is
proportional to $\log\alpha^{-1}$. However, this is not explicitly
exhibited in the current literature; for instance, it can a priori
not be ruled out that terms involving logarithmic corrections
cancel mutually. Moreover, for some models with a mild infrared
behavior \cite{GrHa}, the ground state energy is proven to be
analytic in $\alpha$ (see also \cite{haslerherbst}).

In a recent paper \cite{BCVVii} Chen, Vougalter and the present
authors study the binding energy for Hydrogen atom, which is the
difference between the infimum $\Sigma_0$ of the spectrum of the
translationally invariant operator and the infimum $\Sigma$ of the
spectrum of the operator with Coulomb potential. It is shown in
\cite{BCVVii} that the binding energy as the form
\begin{equation}\label{result-1}
  \Sigma_ 0 - \Sigma\, = \, \frac{\alpha^2}{4}\, +\, e^{(1)} \alpha^3
  \, + \, e^{(2)} \alpha^4 \, + \, e^{(3)}\alpha^5\log\alpha^{-1}
  \, +\, o(\alpha^5\log\alpha^{-1})\, ,
\end{equation}
where the coefficients $e^{(1)}$, $e^{(2)}$ and $e^{(3)}$ are
independent of $\alpha$ and explicitly computed. A natural
question thus arose in the community, to know wether the
logarithmic divergent term in \eqref{result-1} stemmed from
$\Sigma$, $\Sigma_0$ or both. This question can not be answered on
the basis of the computations done in \cite{BCVVii}, because we
did not compute separately the value of $\Sigma$ and $\Sigma_0$,
but their difference.

Although the value of $\Sigma_0$ was known up to the order
$\alpha^3$ from earlier work \cite{BCVVi}, this did not allow us
to answer the above question.

In the work at hand, we compute the infimum $\Sigma_0$ of the
spectrum of the translationally invariant operator, up to the
order $\alpha^4$ with error $\Or(\alpha^5)$, derive $\Sigma$ up to
the order $\alpha^4$, and show that the logarithmic term in
\eqref{result-1} is related to $\Sigma$ and not to $\Sigma_0$.

%%%%%%%%%%%%%%%%%%%%%%%%%%%%%%%%%%%%%%%%%%%%%%%%%%%%%%%%%%%%%%%
%%%%%%%%%%%%%%%%%%%%%%%%%%%%%%%%%%%%%%%%%%%%%%%%%%%%%%%%%%%%%%%
%%%%%%%%%%     THE MODEL                   %%%%%%%%%%%%%%%%%%%%
%%%%%%%%%%%%%%%%%%%%%%%%%%%%%%%%%%%%%%%%%%%%%%%%%%%%%%%%%%%%%%%
%%%%%%%%%%%%%%%%%%%%%%%%%%%%%%%%%%%%%%%%%%%%%%%%%%%%%%%%%%%%%%%
\section{The model}\label{the-model}

%%%%%%%%%%%%%%%%%%%%%%%%%%%%%%%%%%%%%%%%%%%%%%%%%%%%%%%%%%%%%%%
We study a non-relativistic free spinless electron interacting
with the quantized electromagnetic field in Coulomb gauge. The
Hilbert space accounting for the pure states of the electron is
given by $L^2(\R^3)$, where we neglect its spin. The Fock space of
the transverse photons is
 $$
   \gF \; = \; \bigoplus_{n \in \N} \gF_n ,
 $$
where the $n$-photon space $\gF_n =
\bigotimes_s^n\left(L^2(\R^3)\otimes\C^2\right)$ is the symmetric
tensor product of $n$ copies of one-photon Hilbert spaces
$L^2(\R^3)\otimes\C^2$. The factor $\C^2$ accounts for the two
independent transversal polarizations of the photon. On $\gF$, we
introduce creation and annihilation operators $a_\lambda^*(k)$,
$a_\lambda(k)$ satisfying the distributional commutation relations
 $$
  [ \, a_{\lambda}(k) \, , \, a^\ast_{\lambda'}(k') \, ] \; = \;
  \delta_{\lambda, \lambda'} \, \delta (k-k')
  \; \;   ,
  \quad [ \, a_\lambda^\sharp(k) \, ,
  \, a_{\lambda'}^\sharp(k') \, ] \; = \; 0 ,
 $$
where $a^\sharp_\lambda$ denotes either $a_\lambda$ or
$a_\lambda^*$. There exists a unique unit ray $\vac\in\gF$, the
Fock vacuum, which satisfies $a_\lambda(k) \, \vac=0$ for all
$k\in\R^3$ and $\lambda\in\{1,2\}$.

The Hilbert space of states of the system consisting of both the
electron and the radiation field is given by
 $$
  \gH := L^2(\R^3) \otimes \gF\, .
 $$

We shall use units such that $\hbar = c = 1$, and where the mass
of the electron equals $m=1/2$. The electron charge is then given
by $e=\sqrt{\alpha}$.

The Hamiltonian of the system is given by
 $$
  \Htot = I_{el} \otimes H_f \, + \, :\, (i\nabla_x \otimes I_f \,
  - \, \sqrt{\alpha} A(x))^2\, :\, ,
 $$
where $: \, (\cdots) \, :$ denotes normal ordering. The free
photon field energy operator $H_f$ is given by
\begin{equation*}
  H_f = \sum_{\lambda= 1,2} \int_{\R^3} |k| a_\lambda^\ast (k)
  a_\lambda (k) \d k.
\end{equation*}

The magnetic vector potential is
 $$
  A(x) \; = \; \Aa(x) \, + \, \Ac(x) ,
 $$
where
\begin{equation}\label{eq:def-vector-potential}
  \Aa(x) \; = \; \sum_{\lambda=1,2} \int_{\R^3}
  \, \frac{\kappa(|k|)}
  {2\pi |k|^{1/2}} \, \varepsilon_\lambda(k) \, \mathrm{e}^{i k
  x} \, \otimes \, a_\lambda(k) \, \mathrm{d}k
\end{equation}
is the part of $A(x)$ containing the annihilation operators, and
$\Ac(x)=(\Aa(x))^*$. The vectors $\varepsilon_\lambda(k)\in\R^3$
are the two orthonormal polarization vectors perpendicular to $k$,
 $$
  \varepsilon_1(k) = \frac{(k_2, -k_1, 0)}
  {\sqrt{k_1^2 + k_2^2}}\qquad
 {\rm and} \qquad
   \varepsilon_2(k) = \frac{k}{|k|}\wedge \varepsilon_1(k).
 $$
In \eqref{eq:def-vector-potential}, the function $\kappa$
implements an ultraviolet cutoff on the momentum $k$. We assume
$\kappa$ to be of class $C^1$, with compact support in $\{
|k|\leq\Lambda\}$, $0\leq \kappa \leq 1$ and $\kappa = 1$ for
$|k|\leq \Lambda-1$.

The ground state energy of $\Htot$ is denoted by
 $$
  \Sigma_0 : = \inf\mathrm{spec}(\Htot)\, .
 $$

%%%%%%%%%%%%%%%%%%%%%%%%%%%%%%%%%%%%%%%%%%%%%%%%%%%%%%%%%%
%%% def. of self energy operator with total momentum 0 %%%
%%%%%%%%%%%%%%%%%%%%%%%%%%%%%%%%%%%%%%%%%%%%%%%%%%%%%%%%%%

We note that this system is translationally invariant; that is,
$T$ commutes with the operator of total momentum
 $$
 P_{tot} = p_{el}\otimes I_f + I_{el}\otimes P_f ,
 $$
where $p_{el}$ and $P_f$ denote respectively the electron and the
photon momentum operators.

Therefore, for fixed value $p\in\R^3$ of the total momentum, the
restriction of $T$ to the fibre space $\C \otimes\mathfrak{F}$ is
given by (see e.g. \cite{Chen2008})
\begin{equation}\label{def:T(P)}
  T(p) = \ \ :(p - P_f  - \sqrt{\alpha} A(0))^2:  + H_f\, ,
\end{equation}
where by abuse of notation, we dropped all tensor products
involving the identity operators $I_f$ and $I_{el}$. Henceforth,
we will write
 $$
 A^\pm:= A^\pm(0)\ .
 $$

It is proven in \cite{BCFS,Chen2008} that
 $$
 \Sigma_0 = \inf\mathrm{spec} (\H)
 \mbox{ is an eigenvalue of the operator $\H$}\, .
 $$
%%%% end of definition of self energy at momentum P=0 %%%%%%%

%%%%%%%%%%%%   Definitions    %%%%%%%%%%%%%%%%%%%%%%%%%%%%%%

%%%%%%%%%%%%   MAIN THEOREMS   %%%%%%%%%%%%%%%%%%%%%%%%%%%%%%

We are now in position to state our first main result.

On $\gF$ we define respectively the positive bilinear form and its
associated semi-norm
\begin{equation}\label{eq:scalar-product}
 \la\, v\, ,\, w\ra_* := \la\, v,\, (H_f+P_f^2)\, w\, \ra\ ,
 \quad \| v \|_* := \la v,\,
 v\ra_*^\frac12\, .
\end{equation}

%%%%%%%%%%    theorem  ground state energy    %%%%%%%%%%%%%%%
\begin{theorem}[Ground state energy of $\Htot$ and $\H$]\label{thm:main-1}
We have
\begin{equation}\label{eq:main-1}
 \Sigma_0 \ = \
     d^{(0)} \alpha^2\ +\ d^{(1)}\alpha ^3
    \ + d^{(2)} \alpha^4 \  + \ \Or(\alpha^5)\ ,
\end{equation}
with
\begin{equation*}
\begin{split}
    & d^{(0)}  : =
     -\|\ced\|_*^2 \\
    & d^{(1)}  : = 2\| \Aa \ced \|^2
                       - 4\|\cet\|_*^2 - 4\|\ceu\|_*^2 \\
    & d^{(2)}  : =
    - \left(\frac{2\|\Aa\ced\|^2-4\|\ceu\|_*^2 - 4\|\cet\|_*^2
    }{\|\ced\|_*}\right)^2 \\
         & \  + 8 \Re\la \ceu, \Aa\cdot\Aa\cet\ra
    \! + \! 8 \|\Aa\ceu\|^2
    \! + \! 8\|\Aa\cet\|^2
    \! - \! 16 \|\cedd\|_*^2
    \! - \! 16 \|\ceq\|_*^2
    \! + \! \|\ced\|^2 \|\ced\|_*^2 \, ,
\end{split}
\end{equation*}
and
\begin{equation}\label{eq:def-phi}
\begin{split}
\ced  := &  - (H_f+P_f^2)^{-1} \Ac \cdot \Ac \vac\, , \\
 \cet  := &  - (H_f+P_f^2)^{-1} P_f \cdot \Ac \ced\, , \\
 \ceu  := &  - (H_f+P_f^2)^{-1} P_f \cdot \Aa \ced\, , \\
 \cedd := &  - \pr\!\!{}^\perp (H_f+P_f^2)^{-1}
       \left( P_f\cdot\Ac \ceu + P_f\cdot \Aa \cet
       + \frac12\Ac\cdot\Aa\ced\right)
    \\
 \ceq  := &  - (H_f + P_f^2)^{-1} \left(P_f\cdot\Ac\cet + \frac14
 \Ac\cdot\Ac \ced\right)\ ,
\end{split}
\end{equation}
where $\pr\!\!{}^\perp$ is the orthogonal projection onto $\{
\varphi \in\gF\ |\ \la \varphi,\, \ced\ra_*=0\}$.
\end{theorem}
%%%%%%%%%%%%%%%%%%%%%%%%%%%%%%%%%%%%%%%%%%%%%%%%%%%%%%%%%%%%
The proof of Theorem~\ref{thm:main-1} is postponed to
Section~\ref{S-mainproof}. The proof of the upper bound is derived
in subsection~\ref{prf-upper-bound} using a bona fide trial
function, whereas the most difficult part, namely the proof of the
lower bound, is given in subsection~\ref{prf-lower-bound}.

%%%%%%%%%%%%%% corollary on the non analyticity of Sigma%%%%
\begin{corollary}[non analyticity of $\inf\mathrm{spec}(H)$]
The Pauli-Fierz Hamiltonian for an electron interacting with a
Coulomb electrostatic field and coupled to the quantized radiation
field is
 $$
 H:= \Htot - \frac{\alpha}{|x|}\, .
 $$
Its ground state energy $\Sigma:=\inf\,\mathrm{spec}(H)$ fulfills
 $$
  \Sigma =  \tilde{d}^{(0)}\alpha^2
  \, + \, \tilde{d}^{(1)}\alpha^3
  \, + \, \tilde{d}^{(2)}\alpha^4
  \, + \, \tilde{d}^{(3)}\alpha^5\log\alpha^{-1}
  \, + \, o(\alpha^5\log\alpha^{-1})\, ,
 $$
where
 $$
 \tilde{d}^{(0)} = d^{(0)} -\frac14  ,\quad
 \tilde{d}^{(1)} = d^{(1)} - e^{(1)} ,\quad
 \tilde{d}^{(2)} = d^{(2)} - e^{(2)} ,\quad
 \tilde{d}^{(3)} =  - e^{(3)}\, ,
 $$
with
 $$
  e^{(1)} = \frac2{\pi} \int_0^\infty
  \frac{\kappa^2(t)}{1+t} \mathrm{d}t ,
 $$
 \begin{equation*}
 \begin{split}
   & e^{(2)}  =  \frac{2}{3} \, \Re \sum_{i=1}^3 \la (\Aa)^i (H_f+P_f^2)^{-1}\Ac
   . \Ac \vac, (H_f + P_f^2)^{-1} (\Ac)^i \vac \ra
   \\
   & + \frac13
   \sum_{i=1}^3 \|(H_f + P_f^2)^{-\frac12}
   \Big(
   2 \Ac.P_f (H_f + P_f^2)^{-1} (\Ac)^i
   - P_f^i (H_f+P_f^2)^{-1}\Ac. \Ac \Big)\vac\|^2
   \\
   & - \frac23 \sum_{i=1}^3 \| A^- (H_f + P_f^2)^{-1} (A^+)^i
   \vac\|^2
   + 4 a_0^2 \|  Q_1^\perp (-\Delta -\frac{1}{|x|}+\frac14 )^{-\frac12}
  \Delta u_1 \|^2 ,
 \end{split}
 \end{equation*}
\begin{equation*}
  a_0 = \int
  \frac{k_1^2+k_2^2}{4\pi^2|k|^3} \frac{2}{|k|^2
  +|k|} \kappa (|k|)\, \d k_1 \d k_2 \d k_3 ,
\end{equation*}
\begin{equation*}
 e^{(3)} = - \frac{1}{3\pi}
 \| (-\Delta -\frac{1}{|x|} +\frac14)^{\frac12} \nabla
 u_1\|^2 ,
\end{equation*}
and $Q_1^\perp$ is the projection onto the orthogonal complement
to the ground state $u_1$ of the Schr\"odinger operator $-\Delta
-\frac{1}{|x|}$.
\end{corollary}
%%%%%%%%%%%%%% end of corollary on non analyticity  %%%%%%%%
\noindent\textit{Proof.} This is a direct consequence of the above
Theorem~\ref{thm:main-1} and \cite[Theorem~2.1]{BCVVii}.

\medskip

%%%%%%%%%%%%%%%%%%%%%%%%  main result 2  %%%%%%%%%%%%%%%%%%%
The next main result gives an approximate ground state of $\H$.

Let $\gs$ be the ground state of $\H$, normalized by the condition
\begin{equation}\label{eq:norm-1}
 \la\, \gs\, , \, \vac\, \ra = 1\ .
\end{equation}
The existence of $\gs$ was proved in \cite{BCFS, Chen2008}. We
decompose the state $\gs$ according to its $\la\cdot ,
\cdot\ra_*$-projections in the direction of $\ceu$, $\ced$,
$\cedd$, $\cet$ and $\ceq$ and its orthogonal part $R$. This gives
\begin{equation}\label{eq:decomp-1}
  \gs = \vac + 2\eta_1\alpha^\frac32 \ceu + \eta_2 \alpha \ced
  + \tilde\eta_2 \alpha^2\cedd + 2 \eta_3 \alpha^\frac32 \cet
  + \eta_4 \alpha^2 \ceq + R\ ,
\end{equation}
where the coefficients $\eta_i$ ($i=1,2,3,4$) and $\tilde\eta_2$,
and the vector $R$ are uniquely determined by the conditions
\begin{equation}\label{eq:cond-1}
\begin{split}
  & \la \cei, \cej \ra_* = \|\cej\|_*^2 \delta_{ij},\quad
  \la \cei, \cedd \ra_* = 0\, , \\
  & \la \cej,R\ra_* = 0,\quad \la \cedd, R\ra_* =0,\quad
  \la \vac, R\ra = \la\vac,\cej\ra =0\ ,
\end{split}
\end{equation}
for $i,j=1,2,3,4$.
%%%%%%%%%%%%%%  theorem  groud state    %%%%%%%%%%%%%%%%%%%%
\begin{theorem}[Ground state of $\H$]\label{thm:main-2}
Let $\gs$ be the ground state  of $\H$, normalized by the
condition $\la \gs , \vac \ra =1$. Then
\begin{equation}\label{eq:approx-gs}
\begin{split}
 \gs  = & \vac\,  + \,
 2\alpha^\frac32\ceu\,
 + \, \alpha (1-\beta\alpha)\ced\,
 + \, 4\alpha^2 \cedd
 \, + \, 2\alpha^\frac32 \cet\, + \,
 4 \alpha^2 \ceq\, + \, \tilde R\ ,\\
 & \mbox{with }\beta:=  \frac{2\|\Aa \ced\|^2 - 4\|\ceu\|_*^2 - 4\|\cet\|_*^2}
 {\|\ced\|_*^2}\, , \\
 \tilde R := & R \, + \, 2(\eta_1-1)\alpha^\frac32\ceu
 \, + \, (\eta_2 - (1-\beta\alpha))\alpha\ced\,  + \,
 2(\eta_3 - 1)\alpha^\frac32\cet \\
 & \, + \, (\tilde\eta_2 -4)\alpha^2  \cedd
 \, +\,  (\eta_4-4)\alpha^2\ceq\ ,
\end{split}
\end{equation}
where the coefficients $\eta_j$ ($j=1,2,3,4$) and $\tilde\eta_2$
satisfy that there exists a finite constant $c$ such that for all
$\alpha$, $|\eta_{1,3} - 1|^2\leq c\alpha$, $|\eta_2 - 1|^2 \leq
c\alpha^2$, $|\tilde\eta_2-4|^2 \leq c\alpha$, and $|\eta_4 - 4|^2
\leq c \alpha$, and with
 $$
  \|\tilde R\|,\, \|R\| = \Or(\alpha),\quad\mbox{and}\quad
  \|\tilde R\|_*,\, \|R\|_* =\Or(\alpha^2)\, .
 $$
\end{theorem}
The proof of this theorem is given in
subsection~\ref{prf:thm-main-2}.
%%%%%%%%%%%%%%%%%%%%%%%%%%%%%%%%%%%%%%%%%%%%%%%%%%%%%%%%%%%%

%%%%%%%%%%%%%%%%%%%%%%%%%%%%%%%%%%%%%%%%%%%%%%%%
\section{Photon number and field energy bounds}\label{S-photon-bound}
%%%%%%%%%%%%%%%%%%%%%%%%%%%%%%%%%%%%%%%%%%%%%%%%
In order to derive the ground state energy for $\H$, we need to
derive some a priori expected photon number bound and expected
field energy bound for the ground state.

%The first result was derived in \cite{BCVVi}. We recall it for a
%sake of clarity.

%%%%%%%%%%%%%%%%%%%%%%%%%%%%%%%%%%%%%%%%%%%%%%%%%%%%%%%
%\begin{proposition}\label{prop:bound-1}
%The expected photon number in the ground state $\Psi$ of $\H$ is
%bounded by
%  $$
%    \la\, \gs,\, N_f\, \gs\,\ra \leq C \alpha^2 \| \gs \|^2 \ ,
%  $$
%  for a positive constant $C<\infty$ independent of $\alpha$, where
% $$
%   N_f = \sum_{\lambda =+,-} \int  a^*_\lambda(k)
%   a_\lambda(k) \d\, k \ .
% $$
%\end{proposition}
%%%%%%%%%%%%%%%%%%%%%%%%%%%%%%%%%%%%%%%%%%%%%%%%%%%%%%%
%For the proof of this result, see Proposition~3.1 in\cite{BCVVi}.

As a consequence of Theorem~3.2 in \cite{BCVVi}, we have the
following bound for the $*$-norm of the remainder $R$ of the
ground state $\gs$, as defined by \eqref{eq:decomp-1}.
\begin{proposition}\label{prop:field-energy-bound}
There exists $c <\infty$ such that
\begin{equation}\label{eq:field-energy-bound}
 \la (H_f + P_f^2) R, R\ra \leq c (1 + |\tilde{\eta_2}|^2
 + |\eta_4|^2)\alpha^4\, .
\end{equation}
\end{proposition}
%%%%%%%%%%%%%%%%%%%%%%%%%%%%%%%%%%%%%%%%%%%%%%%%%%%%%%%
\textit{Proof.} In \cite[Theorem~3.2]{BCVVi} it is shown that for
$\eta_1$, $\eta_2$ and $\eta_3$ given by \eqref{eq:decomp-1} and
\eqref{eq:cond-1}, and for $r:=\gs - \Omega_f -
2\eta_1\alpha^\frac32\ceu - \eta_2\alpha\ced - 2
\eta_3\alpha^\frac32\cet$, we have $\|r\|_* = \Or(\alpha^2)$.
Therefore, using the decomposition \eqref{eq:decomp-1} concludes
the proof. %\qed
%%%%%%%%%%%%%%%%%%%%%%%%%%%%%%%%%%%%%%%%%%%%%%%%%%%%%%%

%%%%%%%%%%%%%%%%%%%%%%%%%%%%%%%%%%%%%%%%%%%%%%%%%%%%%%%
\begin{proposition}\label{prop:bound-2}
Let
   \begin{equation}\label{eq:def-theta}
     \Theta := \gs - \alpha\eta_2 \ced - 2\alpha^\frac32
     \eta_1\ceu - 2 \alpha^\frac32 \eta_3\cet -\vac\ ,
   \end{equation}
where the vectors $\cei$ ($i=1,2,3$) are defined in
\eqref{eq:def-phi} in Theorem~\ref{thm:main-1} and the
coefficients $\eta_i$ ($i=1,2,3$) are given by the decomposition
of $\gs$ according to \eqref{eq:decomp-1} and the conditions
\eqref{eq:cond-1}.

Then
\begin{equation}\label{eq:estimate-N-theta}
 \la \Theta\, ,\, N_f \Theta\, \ra  = \Or(\alpha^3)\, .
\end{equation}
\end{proposition}
%%%%%%%%%%%%%%%%%%%%%%%%%%%%%%%%%%%%%%%%%%%%%%%%%%%%%%%
\textit{Proof.} According to \cite[Theorem~3.2]{BCVVi}, we have
\begin{equation}\label{eq:prop3.2-1}
 \|\Theta\|^2_* = \Or(\alpha^4)\ .
\end{equation}
Now we write
\begin{equation}\label{eq:prop3.2-2}
  \la \Theta,\, N_f\Theta\ra =
  \int_{|k| < \alpha} \| a_\lambda(k) \Theta\|^2 \d\! k
  + \int_{|k| \geq \alpha} \| a_\lambda(k) \Theta\|^2 \d\! k\ .
\end{equation}
The second term in the right hand side of \eqref{eq:prop3.2-2} is
bounded as follows
\begin{equation}\label{eq:prop3.2-3}
   \int_{|k| \geq \alpha} \| a_\lambda(k) \Theta\|^2 \d\! k
 = \int_{|k| \geq \alpha} \frac{1}{|k|}
 |k|\, \| a_\lambda(k) \Theta\|^2 \d\! k \leq \alpha^{-1}
 \|H_f^\frac12 \Theta\|^2 = \Or(\alpha^3)\ ,
\end{equation}
where we used \eqref{eq:prop3.2-1}. For the first term in the
right hand side of \eqref{eq:prop3.2-2}, we write
\begin{equation}\label{eq:prop3.2-4}
\begin{split}
   & \int_{|k| < \alpha} \| a_\lambda(k) \Theta\|^2 \d\! k
 = \int_{|k| < \alpha} \left\| a_\lambda(k)
 \left(\gs -\alpha \eta_2 \ced - 2\alpha^\frac32\eta_1\ceu
 - 2\alpha^\frac32 \eta_3\cet \right)\right\|^2 \d\! k \\
   & \leq 4 \Big(
 \int_{|k| < \alpha} \| a_\lambda(k)\Psi\|^2 \d\! k
 + \alpha^2 |\eta_2|^2 \int_{|k| < \alpha} \| a_\lambda(k)\ced\|^2 \d\!
 k\\
   & \ \ \ \ \ \ \ + \alpha^3 |\eta_1|^2 \int_{|k| < \alpha} \| a_\lambda(k)\ceu\|^2 \d\! k
 + \alpha^3 |\eta_3|^2 \int_{|k| < \alpha} \| a_\lambda(k)\cet\|^2 \d\! k
 \Big)\ .
\end{split}
\end{equation}
Straightforward computations shows that the last three terms in
the right hand side of \eqref{eq:prop3.2-4} are $\Or(\alpha^3)$.
To estimate the first integral in the right hand side of
\eqref{eq:prop3.2-4} we follow the strategy used in the proof of
\cite[Proposition~3.1]{BCVVi} as explained below.

For $\sigma>0$, let $\Htot_\sigma(p)$ denote the fiber Hamiltonian
regularized by an infrared cutoff implemented by replacing the
ultraviolet cutoff function $\kappa$ of
\eqref{eq:def-vector-potential} by a $C^1$ function
$\kappa_\sigma$ with $\kappa_\sigma=\kappa$ on $[\sigma,\infty)$,
$\kappa_\sigma(0)=0$, and $\kappa_\sigma$ monotonically increasing
on $[0,\sigma]$. Then, $E_\sigma(p):=\inf
\mathrm{spec}(\Htot_\sigma(p))$ is a simple eigenvalue with
eigenvector $\Psi_\sigma(p)\in\gF$ \cite{BCFS, Chen2008}. If
$p=0$, one has $\nabla_pE_\sigma(p=0) = 0$ (see \cite{BCFS,
Chen2008}). In Formula (6.11) of \cite{CF2007}, it is shown that
\begin{equation}\label{eq:prop3.2-4a}
 a_\lambda(k) \Psi_\sigma(0) = (A)\ +\ (B)\, ,
\end{equation}
where from (6.12) of \cite{CF2007}, it follows that
\begin{equation}\label{eq:prop3.2-4b}
 \|\, (A)\, \| \leq C(k) |\nabla_p E_\sigma(0)| = 0\, ,
\end{equation}
and that
\begin{equation}\label{eq:prop3.2-4c}
   (B) = -\sqrt\alpha \frac{\kappa_\sigma(|k|)}{|k|^\frac12}\,
 \frac{1}{\Htot_\sigma(k) - E_\sigma(0) + |k|}
 (\Htot_\sigma(0) -E_\sigma(0))
 \epsilon_\lambda(k) \cdot \nabla_p \Psi_\sigma(0)\, ,
\end{equation}
if the electron spin is zero. Thus it follows immediately from
(6.19) in \cite{CF2007} that
 $$
  \| a_\lambda(k) \Psi_\sigma(0) \| \leq
  c\sqrt\alpha \frac{\kappa_\sigma(|k|)}{|k|}
  \left| \frac{1}{m_{ren,\sigma}}-1\right|\,
  \|\Psi_\sigma(0)\| \leq c\alpha
  \frac{\kappa_\sigma(|k|)}{|k|}\, \|\Psi_\sigma(0)\|\, ,
 $$
for spin zero, where $m_{ren,\sigma}$ is the renormalized electron
mass for $p=0$ (see \cite{BCFS, Chen2008}), defined by
\begin{equation}\label{eq:prop3.2-4d}
 \frac{1}{m_{ren,\sigma}} = 1 - 2
 \frac{\nabla_p\Psi_\sigma(0),\, (\Htot_\sigma(0) - E_\sigma(0))
 \nabla_p \Psi_\sigma(0)\ra}{\| \Psi_\sigma(0) \|^2}\ .
\end{equation}
As proved in \cite{BCFS, Chen2008}, $1<m_{ren,\sigma}<1+ c\alpha$
uniformly in $\sigma\geq 0$.

Therefore, one can write
\begin{equation}\label{eq:prop3.2-5}
\begin{split}
  \int_{|k|\ <\alpha} \| a_\lambda(k) \gs\|^2 \d\! k
  & = \lim_{\sigma\rightarrow 0} \int_{|k| <\alpha} \| a_\lambda(k)
  \gs_\sigma\|^2 \d\! k \\
  & \leq \lim_{\sigma\rightarrow 0} \int_{|k| <\alpha} c\, \alpha
  \frac{\kappa_\sigma(|k|)}{|k|} \, \| \gs_\sigma(0)\|^2 \d\! k =
  \Or(\alpha^3)\ ,
\end{split}
\end{equation}
where $\gs = s-\lim_{\sigma\searrow 0}\Psi_\sigma(0)$ (see
\cite{BCFS}). The inequalities \eqref{eq:prop3.2-3} and
\eqref{eq:prop3.2-5}
conclude the proof. %\qed

A straightforward consequence of this result is
%%%%%%%%%%%%%%%%%%%%%%%%%%%%%%%%%%%%%%%%%%%%%%%%%%%%%%%
\begin{corollary}\label{corollary-1}
For $\Theta$ defined as in \eqref{eq:def-theta} by $\Theta = \gs -
\alpha\eta_2 \ced - 2\alpha^\frac32\eta_1\ceu - 2 \alpha^\frac32
\eta_3\cet -\vac$, we have
\begin{equation}\label{eq:bound-2}
 \| \Theta \|^2 = \Or(\alpha^3)\ .
\end{equation}
\end{corollary}

%%%%%%%%%%%%%%%%%%%%%%%%%%%%%%%%%%%%%%%%%%%%%%%%
\section{Proof of Theorem~\ref{thm:main-1}}\label{S-mainproof}
%%%%%%%%%%%%%%%%%%%%%%%%%%%%%%%%%%%%%%%%%%%%%%%%

We introduce the following notations:
\begin{equation}\label{eq:decomp-2}
\begin{split}
   \cedd & = \ceddu \ + \ \ceddd \ + \ \ceddt \\
        & :=  \left(\pr\!\!{}^\perp (H_f+P_f^2)^{-1}P_f\cdot\Ac
       \ceu\right)
       \  + \ \left(\pr\!\!{}^\perp (H_f+P_f^2)^{-1}
       P_f\cdot \Aa \cet\right) \\
        &\ \ \ \ \  + \ \left(\pr\!\!{}^\perp
       (H_f+P_f^2)^{-1}\frac12\Ac\cdot\Aa\ced\right)\, ,
\end{split}
\end{equation}
\begin{equation}\label{eq:decomp-3}
\begin{split}
   \ceq & = \cequ + \ceqd
  := \Big((H_f + P_f^2)^{-1} P_f\cdot\Ac\cet\Big)\
      +\  \Big(\frac14
                (H_f + P_f^2)^{-1} \Ac\cdot\Ac \ced\Big)\, .
\end{split}
\end{equation}
For $n\in\N$ we also define $\Proj^{(n)}$ as the orthogonal
projection onto the $n$-photon space $\gF_n$ of the Fock space
$\gF$, whereas $\Proj^{(\geq n)}$ shall denote the orthogonal
projection onto $\bigoplus_{k\geq n}\gF_k$.

Finally, we set
\begin{equation}\label{eq:decomp-4}
 R_i := \Proj^{(i)}R,\quad\mbox{for }i=1,2,3,4, \quad
 \mbox{and}\quad R_{\geq k} := \Proj^{(\geq k)}R\ .
\end{equation}
\subsection{Proof of the upper bound}\label{prf-upper-bound}
The proof of the upper bound in Theorem~\ref{thm:main-1} is easily
obtained by picking the trial function
\begin{equation}\label{eq:psi-trial}
\begin{split}
    \pst : = & \vac + 2\alpha^\frac32 \ceu
  + \alpha\left(1 - \alpha\, \frac{\, 2\|\Aa\ced\|^2 - 4\|\ceu\|_*^2
  - 4\|\cet\|_*^2}{\|\ced\|_*^2}\right) \ced \\
    &
  + \alpha^2\cedd + 2 \alpha^\frac32 \cet
  + \alpha^2 \ceq
  \, .
\end{split}
\end{equation}
We then compute $\la \pst,\, \H\pst\ra / \|\pst\|^2$. A
straightforward computation yields
\begin{equation*}
\begin{split}
    & \la \pst,\, \H\pst\ra \\
    & =  -\alpha^2\|\ced\|_*^2
     \ +\ \alpha ^3 \Big( 2\| \Aa \ced \|^2
                       - 4\|\cet\|_*^2 - 4\|\ceu\|_*^2\Big) \\
    & + \ \alpha^4 \Big(  8 \Re\la \ceu,\, \Aa\cdot\Aa\cet\ra
     + 8 \|\Aa\ceu\|^2 + 8\|\Aa\cet\|^2 - 16 \|\cedd\|_*^2
     - 16 \|\ceq\|_*^2 \Big)\\
    & - \alpha^4 \left(\frac{-4\|\ceu\|_*^2 - 4\|\cet\|_*^2 +
 2\|\Aa\ced\|^2}{\|\ced\|_*}\right)^2 + \Or(\alpha^5)\, .
\end{split}
\end{equation*}
Since $\|\pst\|^2 = 1 + \alpha^2 \|\ced\|^2 + \Or(\alpha^3)$, we
thus obtain
\begin{equation*}
\begin{split}
    & \inf\mathrm{spec}(\H)  \leq \frac{\la \pst,\,
   \H\pst\ra}{\|\pst\|^2} \\
    & =
    -\alpha^2\|\ced\|_*^2
     \ +\ \alpha ^3 \Big( 2\| \Aa \ced \|^2
                       - 4\|\cet\|_*^2 - 4\|\ceu\|_*^2\Big)
   + \ \alpha^4 \Big(  8 \Re\la \ceu,\, \Aa\cdot\Aa\cet\ra \\
    &  \ \ \ \ + 8 \|\Aa\ceu\|^2 + 8\|\Aa\cet\|^2
    - 16 \|\cedd\|_*^2
     - 16 \|\ceq\|_*^2 + \|\ced\|^2 \|\ced\|_*^2\Big)\\
    & \ \ \ \
   - \alpha^4 \left(\frac{-4\|\ceu\|_*^2 - 4\|\cet\|_*^2 +
     2\|\Aa\ced\|^2}{\|\ced\|_*}\right)^2
    \ + \ \Or(\alpha^5)\, ,
\end{split}
\end{equation*}
which concludes the proof of the upper bound.

\subsection{Proof of the lower bound}\label{prf-lower-bound}

Since
\begin{equation*}
\begin{split}
    & \H = H_f \ +\ : \, ( -P_f - \alpha^\frac12 A(0))^2\, : \\
    & =\! (H_f\!+\!P_f^2)
  \! + \! \alpha^\frac12 (P_f\!\cdot\! \Ac\! +\!\Ac\!\cdot\! P_f)
  \! + \! \alpha^\frac12 (P_f\!\cdot\! \Aa\! +\! \Aa\!\cdot\! P_f)
  \! + \! \alpha (\Ac)^2
  \! + \! \alpha (\Aa)^2 + 2\alpha \Ac\!\cdot\!\Aa
\end{split}
\end{equation*}
we obtain
\begin{equation}\label{eq:proof-1}
\begin{split}
 \la \gs,\, \H \gs\ra
 = &  \Re\la \gs, \, \alpha^\frac12 P_f\cdot\Aa\gs\ra
    + \Re\la \gs, \, 2\alpha\Aa\cdot\Aa\gs\ra \\
 &
    + \la \gs , \, 2\alpha \Ac\cdot\Aa\gs\ra
    + \la \gs , \, (H_f+P_f^2)\gs\ra
\end{split}
\end{equation}
As in \eqref{eq:decomp-1}-\eqref{eq:cond-1}, we decompose the
ground state $\gs$ of $\H$ as follows
 $$
  \gs = \vac + 2\eta_1\alpha^\frac32 \ceu + \eta_2 \alpha \ced
  + \tilde\eta_2 \alpha^2\cedd + 2 \eta_3 \alpha^\frac32 \cet
  + \eta_4 \alpha^2 \ceq + R\ .
 $$

Each term in the right hand side of \eqref{eq:proof-1} are
estimated respectively in
Lemmata~\ref{lem:appendix-1}-\ref{lem:appendix-4}.

We thus collect all terms that occur in
Lemmata~\ref{lem:appendix-1}-\ref{lem:appendix-4}, regroup them
according to the following rearrangement, and estimate them
separately
\begin{equation}\label{eq:splitting}
\begin{split}
 \la \gs,\, \H \gs\ra = (I) + (II) + (III) + (IV) + (V)
 + \ \mbox{positive terms} \,
\end{split}
\end{equation}
where the positive terms are a part of $\la \gs,
(H_f+P_f^2)\gs\ra$ and
\begin{eqnarray*}
 (I)   & = & \mbox{Terms with a pre-factor $\alpha^2$ involving a
               remainder term $R_i$,} \\
 (II)  & = & \mbox{Terms with a pre-factor $\alpha^2$ not involving
               remainder terms $R$,}\\
 (III) & = & \mbox{Terms with a pre-factor $\alpha^3$,}\\
 (IV)  & = & \mbox{Terms with a prefactor $\alpha^4$,}\\
 (V)   & = & \mbox{Terms with a pre-factor $\alpha^5$ and the terms
               $\Or(\alpha^5)$.}
\end{eqnarray*}

$\bullet$ Terms with a pre-factor $\alpha^2$ involving a remainder
term $R_i$.
\begin{equation*}
\begin{split}
   (I) := & \
 - 8 \alpha^2 \Re\eta_1 \la \ceddu,\, R_2\ra_*
 - 8 \alpha^2 \Re\eta_3 \la \ceddd,\, R_2\ra_*
 - 8\alpha^2  \Re\eta_2 \la \ceddt,\, R_2\ra_* \\
   &
 - 8 \alpha^2 \Re\eta_2 \la \ceqd,\,  R_4\ra_*
 - 8 \alpha^2 \Re\eta_3 \la \cequ,\,  R_4\ra_*
\end{split}
\end{equation*}

$\bullet$ Terms with a pre-factor $\alpha^2$ not involving
remainder terms $R$:
\begin{equation}\label{eq:alpha2-1}
\begin{split}
   (II) & : = -\alpha^2 2\Re \bar\eta_2 \|\ced\|_*^2
 + \alpha^2 |\eta_2|^2 \|\ced\|_*^2 \\
   & = -\alpha^2 \|\ced\|_*^2 + \alpha^2 \left( (\Re\eta_2) -1)^2 +
 (\Im \eta_2)^2 \right) \\
   &
 \geq
 -\alpha^2 \|\ced\|_*^2 + \alpha^2 \left( (\Re\eta_2) -1)^2
 \right)\, .
\end{split}
\end{equation}

$\bullet$ Terms with a pre-factor $\alpha^3$.
\begin{equation}\label{eq:alpha3-1}
\begin{split}
   (III) &:=
 \alpha^3\Big(\!-\! 8\Re \eta_1\bar\eta_2 \|\ceu\|_*^2
 \!+\! 4 |\eta_1|^2 \|\ceu\|_*^2 \\
   &  - 8  \Re \eta_3\bar\eta_2 \|\cet\|_*^2
 \!+\! 4  |\eta_3|^2 \|\cet\|_*^2
 + 2 |\eta_2|  \|\Aa\ced\|^2\Big)\\
   & =
 4\alpha^3 \|\ceu\|_*^2 \left( |\eta_1-\eta_2|^2 -|\eta_2|^2\right)\\
   & + 4\alpha^3 \|\cet\|_*^2 \left( |\eta_3-\eta_2|^2 -
 |\eta_2|^2\right) + 2\alpha^3\|\Aa \ced\|^2 + \Or(\alpha^5)\, .
\end{split}
\end{equation}
Since from Lemma~\ref{lem:bcvv1} we have $\eta_2= 1+\Or(\alpha)$,
we get $(\Im\eta_2)^2=\Or(\alpha^2)$ and $(\Re\eta_2)^2 - 1 =2
(\Re\eta_2 -1) + \Or(\alpha^2)$, and thus $|\eta_2|^2 = 1+
2(\Re\eta_2 -1) + \Or(\alpha^2)$. Together with
\eqref{eq:alpha3-1}, this yields
\begin{equation}\label{eq:alpha3-2}
\begin{split}
 (III) = &
 \alpha^3 (-4\|\ceu\|_*^2 - 4\|\cet\|_*^2 + 2\|\Aa\ced\|^2) \\
 & + 2\alpha^3 (\Re\eta_2 - 1)
 (-4\|\ceu\|_*^2 - 4\|\cet\|_*^2 + 2\|\Aa\ced\|^2) \\
 & + 4\alpha^3\|\ceu\|_*^2 |\eta_1 - \eta_2|^2 +
 4\alpha^3\|\cet\|_*^2 |\eta_3-\eta_2|^2 + \Or(\alpha^5)\, .
\end{split}
\end{equation}
The first term in the right hand side of \eqref{eq:alpha3-2} is
the $\alpha^3$ term in the equality \eqref{eq:main-1}, thus we
leave it as it is. The last line in \eqref{eq:alpha3-2}, which is
positive, shall be used later to estimate the terms $(I)$ and
$(IV)$.

The second term in the right hand side of \eqref{eq:alpha3-2} is
estimated together with the term $\alpha^2 \|\ceu\|_*^2
(\Re\eta_2-1)^2$ obtained in the lower bound \eqref{eq:alpha2-1}
for $(II)$. We obtain
\begin{equation}\label{eq:alpha3-3}
\begin{split}
 & 2 \alpha^3 (\Re\eta_2 -1)
 (2\|\Aa\ced\|^2 - 4\|\ceu\|_*^2 - 4\|\cet\|_*^2)
 + \alpha^2 \|\ced\|_*^2 (\Re\eta_2 -1)^2 \\
 & = \alpha^2
 \left(\alpha
 \frac{2\|\Aa\ced\|^2 -4\|\ceu\|_*^2 - 4\|\cet\|_*^2}{\|\ced\|_*}
 + (\Re\eta_2 -1)\|\ced\|_*\right)^2 \\
 & \ \ \ \ - \alpha^4
 \left(
 \frac{2\|\Aa\ced\|^2 -4\|\ceu\|_*^2 - 4\|\cet\|_*^2}
 {\|\ced\|_*}\right)^2 \\
 & \geq
 - \alpha^4 \left(\frac{2\|\Aa\ced\|^2
   -4 \|\ceu\|_*^2 - 4\|\cet\|_*^2}{\|\ced\|_*}\right)^2 \, .
\end{split}
\end{equation}

$\bullet$ Collecting estimates \eqref{eq:alpha2-1},
\eqref{eq:alpha3-2} and \eqref{eq:alpha3-3} yields
\begin{equation}\label{eq:II+III}
\begin{split}
 (II) + (III) \geq
 & -\alpha^2\|\ced\|_*^2 + \alpha^3
 \Big( 2\|\Aa\ced\|^2 -4\|\ceu\|_*^2
 - 4\|\cet\|_*^2 \Big) \\
 & - \alpha^4 \left(\frac{2\|\Aa\ced\|^2
  - 4\|\ceu\|_*^2 - 4\|\cet\|_*^2}{\|\ced\|_*}\right)^2 \\
 & + 4\alpha^3\|\ceu\|_*^2 |\eta_1 - \eta_2|^2 +
 4\alpha^3\|\cet\|_*^2 |\eta_3-\eta_2|^2 +\Or(\alpha^5)\, .
\end{split}
\end{equation}

$\bullet$ Terms with a pre-factor $\alpha^4$.
\begin{equation}\label{eq:alpha4-1}
\begin{split}
   & (IV) := \\
   & - 8\alpha^4
  \Big(\Re \eta_1 \bar{\tilde{\eta_2}} \la \ceddu, \cedd \ra_*
      + \Re \eta_3 \bar{\tilde{\eta_2}} \la \ceddd, \cedd \ra_*
      + \Re \eta_2 \bar{\tilde{\eta_2}} \la \ceddt, \cedd \ra_*\Big)
      \!+\! |\tilde\eta_2|^2 \alpha^4 \|\cedd\|_*^2\\
   & - 8 \alpha^4\Big( \Re\eta_2\bar\eta_4 \la\ceqd,\,
  \ceq\ra_*
  + \Re\eta_3\bar\eta_4 \la\cequ,\,
  \ceq\ra_*\Big) + |\eta_4|^2 \alpha^4 \|\ceq\|_*^2\\
   & + 8 \alpha^4 \Re\eta_1\bar\eta_3 \la \ceu,\,
  \Aa\cdot\Aa\cet\ra
  + 8 |\eta_1|^2 \alpha^4 \|\Aa\ceu\|^2 + 8 |\eta_3|^2 \alpha^4
  \|\Aa\cet\|^2\ .
\end{split}
\end{equation}
Let us first remark that in this expression, we have terms with
pre-factor $\tilde\eta_2$ and positive terms with pre-factor
$|\tilde\eta_2|^2$, therefore, this implies that $\tilde\eta_2$ is
uniformly bounded in $\alpha$ for a minimizer.  The same remarks
hold for $\eta_4$. Thus, there exists $c<\infty$ independent on
$\alpha$ such that
\begin{equation}\label{eq:eta-bounds}
  |\tilde\eta_2| \leq c,\quad |\eta_4|\leq c\, .
\end{equation}

Now, we add to the term $(IV)$ half of the positive term
$4\alpha^3\|\ceu\|_*^2 |\eta_1 - \eta_2|^2 + 4\alpha^3\|\cet\|_*^2
|\eta_3-\eta_2|^2$ obtained in the lower bound \eqref{eq:II+III}
for $(II)+(III)$, and we split the resulting expression in three
parts as follows
\begin{equation}\label{eq:split-IV}
\begin{split}
 & (IV) + 2\alpha^3\| \ceu \|_*^2 |\eta_1 - \eta_2|^2 +
 2\alpha^3\| \cet \|_*^2 |\eta_3-\eta_2|^2 \\
 & =: (IV)^{(1)}
 + (IV)^{(2)} + (IV)^{(3)} + (IV)^{(4)}\, ,
\end{split}
\end{equation}
where
\begin{equation}\label{eq:def-IV1}
\begin{split}
       (IV)^{(1)} := & - 8\alpha^4
  \Big(\Re \eta_1 \bar{\tilde{\eta_2}} \la \ceddu, \cedd \ra_*
      + \Re \eta_3 \bar{\tilde{\eta_2}} \la \ceddd, \cedd \ra_*
      + \Re \eta_2 \bar{\tilde{\eta_2}} \la \ceddt, \cedd \ra_*\Big)\\
       & + |\tilde\eta_2|^2 \alpha^4 \|\cedd\|_*^2
      + 2 \alpha^3\| \ceu \|_*^2 |\eta_1 - \eta_2|^2 +
         \alpha^3\| \cet \|_*^2 |\eta_3-\eta_2|^2
   \, ,%
\end{split}
\end{equation}
\begin{equation}\label{eq:def-IV2}
\begin{split}
     (IV)^{(2)}:= &
  \  - 8 \alpha^4\Big( \Re\eta_2\bar\eta_4 \la\ceqd,\,
  \ceq\ra_*
  + \Re\eta_3\bar\eta_4 \la\cequ,\,
  \ceq\ra_*\Big) + |\eta_4|^2 \alpha^4 \|\ceq\|_*^2 \\
     & +  \alpha^3\| \cet \|_*^2 |\eta_3-\eta_2|^2 \, ,
\end{split}
\end{equation}
and
\begin{equation}\label{eq:def-IV3}
\begin{split}
    (IV)^{(3)} :=  & \  8 \alpha^4 \Re\eta_1\bar\eta_3 \la \ceu,\,
  \Aa\cdot\Aa\cet\ra
  + 8 |\eta_1|^2 \alpha^4 \|\Aa\ceu\|^2\\
    & + 8 |\eta_3|^2 \alpha^4
  \|\Aa\cet\|^2\
      + \alpha^3\| \ceu \|_*^2 |\eta_1 - \eta_2|^2 +
      \alpha^3\| \cet \|_*^2 |\eta_3-\eta_2|^2
\end{split}
\end{equation}

Using from Lemma~\ref{lem:bcvv1} that $\eta_2 = 1 +\Or(\alpha)$
and the fact that $\eta_4$ is bounded uniformly in $\alpha$ (see
\eqref{eq:eta-bounds}) yields
\begin{equation}\label{eq:IV2-1}
\begin{split}
   & (IV)^{(2)} =
    -8 \alpha^4 \left( \Re \eta_2 \bar\eta_4 \la \ceqd, \ceq\ra_*
   + \Re \eta_2 \bar \eta_4 \la \cequ,\ceq\ra_* \right)\\
   & - 8 \alpha^4 \Re (\eta_3-\eta_2) \bar\eta_4 \la \cequ,\ceq\ra_*
   + \alpha^3 \|\cet\|_*^2 |\eta_3 - \eta_2|^2
   + |\eta_4|^2 \alpha^4 \|\ceq\|_*^2  \\
     & = -8 \alpha^4 \left( \Re \eta_2 \bar\eta_4 \la \ceqd, \ceq\ra_*
   + \Re \eta_2 \bar \eta_4 \la \cequ,\ceq\ra_* \right)
   - c \alpha^5 |\eta_4|^2
   + |\eta_4|^2 \alpha^4 \|\ceq\|_*^2 \\
   & \geq - 16 \alpha^4 \|\ceq\|_*^2 - c\alpha^5 \, .
\end{split}
\end{equation}

The term $(IV)^{(1)}$ is treated as follows
\begin{equation}\label{eq:IV1-1}
\begin{split}
    (IV)^{(1)} =
 & - 8\alpha^4 \Re (\eta_1-\eta_2)
 \bar{\tilde{\eta_2}} \la \ceddu, \cedd\ra_*
 + 2 \alpha^3\| \ceu \|_*^2 |\eta_1 - \eta_2|^2 \\
   &
 - 8 \alpha^4 \Re (\eta_3-\eta_2) \bar{\tilde{\eta_2}}
   \la \ceddd, \cedd \ra_*
   + \alpha^3\| \cet \|_*^2 |\eta_3 - \eta_2|^2 \\
   & - 8\alpha^4 \Re \eta_2
 \bar{\tilde{\eta_2}} \la \ceddu, \cedd\ra_*
 - 8 \alpha^4 \Re \eta_2 \bar{\tilde{\eta_2}}
   \la \ceddd, \cedd \ra_*
 - 8\alpha^4 \Re \eta_2 \bar{\tilde{\eta_2}}\la \ceddt,\cedd\ra_*
   \\
   & + |\tilde\eta_2|^2 \alpha^4 \|\cedd\|_*^2 \, .
\end{split}
\end{equation}
Since $\tilde\eta_2$ is bounded (see \eqref{eq:eta-bounds}), the
first line and the second line in the right hand side are of the
order $\alpha^5$. In addition, replacing $\eta_2$ by
$1+\Or(\alpha)$ (see Lemma~\ref{lem:bcvv1}) in the third line of
\eqref{eq:IV1-1} yields
\begin{equation}\label{eq:IV1-2}
\begin{split}
   & (IV)^{(1)} = \\
   & - 8\alpha^4 \Re\Big(
   \bar{\tilde{\eta_2}} \la \ceddu, \cedd\ra_*
 + \bar{\tilde{\eta_2}} \la \ceddd, \cedd \ra_*
 + \bar{\tilde{\eta_2}} \la \ceddt,\cedd\ra_*\Big)
 + |\tilde\eta_2|^2 \alpha^4 \|\cedd\|_*^2 +\Or(\alpha^5)\\
   & = - 8\alpha^4\Re\tilde{\eta_2} \|\cedd\|_*^2
 + |\tilde\eta_2|^2 \alpha^4 \|\cedd\|_*^2 +\Or(\alpha^5) \\
   & \geq -16 \alpha^4 \|\cedd\|_*^2 + \Or(\alpha^5)\, .
\end{split}
\end{equation}

Eventually, we estimate the term $(IV)^{(3)}$. We have
\begin{equation}\label{eq:IV3-1}
\begin{split}
   (IV)^{(3)} =  & 8 \alpha^4 \Re\eta_1\bar\eta_3 \la \ceu,\,
  \Aa\cdot\Aa\cet\ra
  + \frac12 \alpha^3\| \ceu \|_*^2 |\eta_1 - \eta_2|^2
  + \frac12 \alpha^3\| \cet \|_*^2 |\eta_3-\eta_2|^2 \\
   & + 8 |\eta_1|^2 \alpha^4 \|\Aa\ceu\|^2
    + \frac12 \alpha^3\| \ceu \|_*^2 |\eta_1 - \eta_2|^2 \\
   & + 8 |\eta_3|^2 \alpha^4 \|\Aa\cet\|^2\
    + \frac12 \alpha^3\| \cet \|_*^2 |\eta_3-\eta_2|^2
\end{split}
\end{equation}
The first line in \eqref{eq:IV3-1} is estimated as
\begin{equation}\label{eq:IV3-2}
\begin{split}
   & 8 \alpha^4 \Re\eta_1\bar\eta_3 \la \ceu,\,
  \Aa\cdot\Aa\cet\ra
  + \frac12 \alpha^3\| \ceu \|_*^2 |\eta_1 - \eta_2|^2
  + \frac12 \alpha^3\| \cet \|_*^2 |\eta_3-\eta_2|^2 \\
   & =
    8 \alpha^4 \Re(\eta_1-\eta_2)\bar\eta_3 \la \ceu,\,
  \Aa\cdot\Aa\cet\ra
  + \frac12 \alpha^3\| \ceu \|_*^2 |\eta_1 - \eta_2|^2 \\
   &
    + 8 \alpha^4 \Re\eta_2(\bar\eta_3 -\bar\eta_2)\la \ceu,\,
  \Aa\cdot\Aa\cet\ra
  + \frac12 \alpha^3\| \cet \|_*^2 |\eta_3-\eta_2|^2\\
   & + 8 \alpha^4 \Re |\eta_2|^2 \la \ceu,\,
  \Aa\cdot\Aa\cet\ra \\
    & \geq -c\alpha^5  + 8 \alpha^4\Re \la
  \ceu,\Aa\cdot\Aa\cet\ra \,
\end{split}
\end{equation}
where we used again $|\eta_2|^2 = 1+\Or(\alpha)$ and $|\eta_3| =
\Or(1)$.

The second line in \eqref{eq:IV3-1} is estimated as
\begin{equation}\label{eq:IV3-3}
\begin{split}
    & 8 |\eta_1|^2 \alpha^4 \|\Aa\ceu\|^2
    + \frac12 \alpha^3\| \ceu \|_*^2 |\eta_1 - \eta_2|^2 \\
    & = 8 |\eta_2|^2 \alpha^4 \|\Aa\ceu\|^2
    + 8 (|\eta_1|^2-|\eta_2|^2) \alpha^4 \|\Aa\ceu\|^2
    + \frac12 \alpha^3\| \ceu \|_*^2 |\eta_1 - \eta_2|^2 \\
     & \geq
       8  \alpha^4 \|\Aa\ceu\|^2 + \Or(\alpha^5)
    - 96 |\eta_1 - \eta_2|\alpha^4 \|\Aa\ceu\|^2
      + \frac12 \alpha^3\| \ceu \|_*^2 |\eta_1 - \eta_2|^2 \\
    & \geq 8  \alpha^4 \|\Aa\ceu\|^2 + \Or(\alpha^5)\, .
\end{split}
\end{equation}

Similarly, the third line in \eqref{eq:IV3-1} is estimated by
\begin{equation}\label{eq:IV3-4}
\begin{split}
    & 8 |\eta_3|^2 \alpha^4 \|\Aa\cet\|^2\
    + \frac12 \alpha^3\| \cet \|_*^2 |\eta_3-\eta_2|^2
  \geq  8 \alpha^4 \|\Aa\cet\|^2 + \Or(\alpha^5)\, .
\end{split}
\end{equation}

Collecting \eqref{eq:IV3-2}, \eqref{eq:IV3-3} and \eqref{eq:IV3-4}
yields
\begin{equation}\label{eq:IV3-5}
\begin{split}
   (IV)^{(3)} \geq
  8 \alpha^4\Re \la
  \ceu,\Aa\cdot\Aa\cet\ra
  + 8  \alpha^4 \|\Aa\ceu\|^2
  + 8  \alpha^4 \|\Aa\cet\|^2
  + \Or(\alpha^5)\, .
\end{split}
\end{equation}

This inequality, together with \eqref{eq:split-IV},
\eqref{eq:IV2-1} and \eqref{eq:IV1-2} gives
\begin{equation}\label{eq:IV-final}
\begin{split}
   & (IV) \, + \, 2\alpha^3\| \ceu \|_*^2 |\eta_1 - \eta_2|^2
 \, + \, 2\alpha^3\| \cet \|_*^2 |\eta_3-\eta_2|^2 \\
   & \geq \alpha^4\Big(
  8 \Re \la
  \ceu,\Aa\cdot\Aa\cet\ra
  + 8   \|\Aa\ceu\|^2
  + 8   \|\Aa\cet\|^2
  - 16  \|\ced\|_*^2
  - 16  \|\ceq\|_*^2 \Big) \\
    &
  \ \ \ + \Or(\alpha^5)\, .
\end{split}
\end{equation}

$\bullet$ Next, we can treat the term $(I)$. For that sake, we add
the remaining other half of the positive term
$4\alpha^3\|\ceu\|_*^2 |\eta_1 - \eta_2|^2 + 4\alpha^3\|\cet\|_*^2
|\eta_3-\eta_2|^2$ obtained in the lower bound \eqref{eq:II+III}
for $(II)+(III)$. Writing $\eta_1 = (\eta_1-\eta_2) + \eta_2$,
$\eta_3 = (\eta_3-\eta_2) + \eta_2$, $ \eta_2 = 1+\Or(\alpha)$ and
using the fact that $\la \ceddu, R_2\ra_* + \la\ceddd, R_2\ra_* +
\la \ceddt,R_2\ra_* = \la \cedd,R_2\ra_*=0$, we get, following the
same arguments as for the estimate of $(IV)$
\begin{equation}\label{eq:I-1}
\begin{split}
 (I) + 2\alpha^3\|\ceu\|_*^2 |\eta_1 - \eta_2|^2
     + 2\alpha^3\|\cet\|_*^2 |\eta_3-\eta_2|^2
 =\Or(\alpha^5)\, .
\end{split}
\end{equation}

$\bullet$ Terms with a pre-factor $\alpha^5$ and the terms
$\Or(\alpha^5)$.

Collecting these terms yields the following result
\begin{equation}\label{eq:alpha5-1}
\begin{split}
  (V) := -c \alpha^5(|\eta_1|^2 + |\eta_2|^2 + |\eta_3|^2 + |\eta_4|^2 +
  |\tilde\eta_2|^2 ) + \Or(\alpha^5) = \Or(\alpha^5)\, .
\end{split}
\end{equation}
The last equality holds since $\eta_1$, $\eta_2$, and $\eta_3$ are
bounded (Lemma~\ref{lem:bcvv1}) and since we proved in
\eqref{eq:eta-bounds} that $\tilde\eta_2$ and $\eta_4$ are also
bounded.

$\bullet$ Collecting \eqref{eq:II+III}, \eqref{eq:IV-final}
\eqref{eq:I-1} and \eqref{eq:alpha5-1} thus gives
\begin{equation}\label{eq:collect-1}
\begin{split}
   & \la \gs,\, \H \gs \ra =
 (I) + (II) + (III) + (IV) + (V) \\
   & \geq
 -\alpha^2\|\ced\|_*^2 + \alpha^3
 \Big( 2\|\Aa\ced\|^2 -4\|\ceu\|_*^2
 - 4\|\cet\|_*^2 \Big) \\
   &  - \alpha^4 \left(\frac{-4\|\ceu\|_*^2 - 4\|\cet\|_*^2 +
 2\|\Aa\ced\|^2}{\|\ced\|_*}\right)^2 \\
   &  + \alpha^4\Big(
  8 \Re \la
  \ceu,\Aa\cdot\Aa\cet\ra
  \! + \! 8   \|\Aa\ceu\|^2
  \! + \! 8   \|\Aa\cet\|^2
  \! - \! 16  \|\ced\|_*^2
  \! - \! 16  \|\ceq\|_*^2 \Big)
  \!+\! \Or(\alpha^5)\, .
\end{split}
\end{equation}

We conclude the proof of the lower bound for
$\inf\mathrm{spec}(\H))$ by computing
\begin{equation}\label{eq:norm}
\begin{split}
  \| \gs\|^2  = &
  1 + \|2\eta_1\alpha^\frac32\ceu + R_1\|^2
  + \|\eta_2\alpha\ced + \tilde\eta_2\alpha^2\cedd + R_2\|^2\\
  & + \| 2\eta_3\alpha^\frac32\cet +R_3\|^2 + \|\alpha^2\eta_4\ceq +
  R_4\|^2 \\
  = & 1 + \alpha^2 \|\ced\|^2  + \Or(\alpha^3)\ ,
\end{split}
\end{equation}
where we used that $\eta_1$, $\eta_3$, $\tilde{\eta_2}$, and
$\eta_4$ are bounded (Lemma~\ref{lem:bcvv1} and
\eqref{eq:eta-bounds}), that $\eta_2 = 1+\Or(\alpha)$
(Lemma~\ref{lem:bcvv1}), and as a consequence of
Corollary~\ref{corollary-1} that the following holds: $\|R_1\|,\,
\|R_2\|,\, \|R_3\|, \|R_{\geq 4}\| =\Or(\alpha^\frac32)$.
%%%%%%%%%%%%%%%%%%%%%%%%%%%%%%%%%%%%%%%%%%%%%%%%

\subsection{Proof of Theorem~\ref{thm:main-2}}\label{prf:thm-main-2}
The proof of \eqref{eq:approx-gs} is a consequence of the fact
that $\inf\mathrm{spec}(\H) = \la\gs,\, \H\gs\ra/\|\gs\|^2$ and
the value of $\la\pst,\, \H\pst\ra/\|\pst\|^2$ coincide up to
$\Or(\alpha^5)$ for $\pst : = \vac + 2\alpha^\frac32 \ceu +
\alpha\left(1 - \alpha\, \frac{\, 2\|\Aa\ced\|^2 - 4\|\ceu\|_*^2 -
4\|\cet\|_*^2}{\|\ced\|_*^2} \right) \ced + \alpha^2\cedd + 2
\alpha^\frac32 \cet + \alpha^2 \ceq$.

The properties for $\eta_1,3$ and $\eta_2$ were already
established in \cite{BCVVi} as reminded in Lemma~\ref{lem:bcvv1}.
The properties for $\tilde\eta_2$ and $\eta_4$ come from the fact
that $\tilde\eta_2$ and $\eta_4$ minimize \eqref{eq:collect-1} up
to $\Or(\alpha^5)$.

The equality $\|R\|_*=\Or(\alpha^2)$ is given by
Proposition~\ref{prop:field-energy-bound}. The equality $\|\tilde
R\|_* = \Or(\alpha^2)$ is a consequence of $\| R
\|_*=\Or(\alpha^2)$, the definition \eqref{eq:approx-gs} for
$\tilde R$, and the $*$-orthogonalities in \eqref{eq:cond-1}.

Finally, Corollary~\ref{corollary-1} proves $\| R \|=\Or(\alpha)$,
which in turn implies $\|\tilde R\|=\Or(\alpha)$. %\qed

%%%%%%%%%%%%%%%%%%%%%%%%%%%%%%%%%%%%%%%%%%%%%%
%%%%%%%%%%%  ACKNOWLEDGEMENTS %%%%%%%%%%%%%%%%
%%%%%%%%%%%%%%%%%%%%%%%%%%%%%%%%%%%%%%%%%%%%%%
\noindent\textbf{Acknowledgments}

J.-M.~B. and S.~A.~V. thank the Institute for Mathematical
Sciences and the Centre of Quantum Technologies of the National
University of Singapore, where this work was done. J.-M. B. also
gratefully acknowledges financial support from Agence Nationale de
la Recherche, via the project HAM-MARK ANR-09-BLAN-0098-01.

%%%%%%%%%%%%%%%%%%%%%%%%%%%%%%%%%%%%%%%%%%%%%%%%
%\begin{appendix}\label{appendix}
\appendix
\section*{Appendix}\label{appendix}
\setcounter{section}{1}
\subsection{Estimates on $\eta_1$, $\eta_2$ and $\eta_3$}
In the following lemma, we give an estimate of the coefficients
$\eta_1$, $\eta_2$ and $\eta_3$ that occur in the decomposition
\eqref{eq:decomp-1} of $\gs$.
\begin{lemma}\label{lem:bcvv1}
We have
\begin{equation}\label{eq:estimate-eta-i}
 \eta_1 = 1 + \Or(\alpha^\frac12),\quad \eta_3 = 1 +
 \Or(\alpha^\frac12),\quad\mbox{and}\quad \eta_2 = 1 + \Or(\alpha)
\end{equation}
\end{lemma}
\textit{Proof.} This is a direct consequence on the estimates of
$\eta_1$, $\eta_2$ and $\eta_3$ for the approximate ground state
up to the order $\alpha^3$ derived in \cite{BCVVi}, since, due to
the conditions \eqref{eq:cond-1}, the coefficients $\eta_1$,
$\eta_2$ and $\eta_3$ in the decomposition \eqref{eq:decomp-1} of
$\gs$ are the same as the coefficients $\eta_1$, $\eta_2$ and
$\eta_3$ in the decomposition \cite[(10)]{BCVVi}. Note that there
was a misprint in the estimates provided in \cite{BCVVi} for
$|\eta_1-1|$, $|\eta_2-1|$ and $|\eta_3 -1|$, since a square was
missing. One should read in \cite[Theorem~3.2]{BCVVi},
$|\eta_{1,3}-1|^2 \leq c \alpha$ and $|\eta_2-1|^2 \leq c
\alpha^2$.
%%%%%%%%%%%%%%%%%%%%%%%%%%%%%%%%%%%%%%%%%%%%%%%%
\subsection{Estimate of the term
$\Re\la\gs,\, 4 \alpha^\frac12 P_f\cdot\Aa\gs\ra$}

%\begin{lemma}\label{rem:A}
%For or all integer $n$, for all $((k_1, \lambda_1), \
%(k_2,\lambda_2), \ldots,\ (k_n, \lambda_n))\in
%\left((\R^3\setminus\{k, \ |k|\geq\Lambda+1\})\times\{1,2\}
%\right)^n$, we have
%\begin{equation}\label{eq:cutoff}
%  \Proj^n\gs_{\leq \Lambda}((k_1, \lambda_1),
%  \ (k_2,\lambda_2), \ldots,\ (k_n, \lambda_n)) = 0
%\end{equation}
%and
% $$
% \langle \Phi_{\leq \Lambda}, \H \Phi_{\leq\Lambda} \rangle
% \leq \langle \Phi, \H \Phi\rangle\, .
% $$
%\end{lemma}
%\noindent\textit{Proof.} We have
% $$
%  \H = P_f^2 + 2\sqrt{\alpha}\Re P_f \cdot A(0)
%  + \alpha :A(0)^2: + H_f \, ,
% $$
%and there is a cutoff function $\kappa(|k|)$ in the definition of
%$A(0)$. Therefore, for any given normalized state $\Phi\in\gH$,
%there exists a normalized state $\Phi_{\leq\Lambda}\in\gF$ such
%that, for all $n\in\{1,2,\ldots\}$, for all $((k_1, \lambda_1), \
%(k_2,\lambda_2), \ldots,\ (k_n, \lambda_n))\in
%\left((\R^3\setminus\{k, \ |k|\geq\Lambda+1\})\times\{1,2\}
%\right)^n$, we have
%\begin{equation}\label{eq:cutoff}
%  \Proj^n\Phi_{\leq \Lambda}((k_1, \lambda_1),
%  \ (k_2,\lambda_2), \ldots,\ (k_n, \lambda_n)) = 0
%\end{equation}
%and
% $$
% \langle \Phi_{\leq \Lambda}, \H \Phi_{\leq\Lambda} \rangle
% \leq \langle \Phi, \H \Phi\rangle\, .
% $$
%This concludes the proof.

Throughout this appendix, we shall always use the decomposition of
$\gs$ given by \eqref{eq:decomp-1}-\eqref{eq:cond-1} and
\eqref{eq:decomp-2}-\eqref{eq:decomp-3}.

%%%%%%%%%%%%%%%%%%%%%%%%%%%%%%%%%%%%%%%%%%%%%%%%
\begin{lemma}\label{lem:appendix-1}
We have
\begin{equation}\label{eq:main-appendix-1}
\begin{split}
   & \Re\la \gs ,\, 4 \alpha^\frac12 P_f\cdot\Aa\gs\ra
 + \frac{1}{8} \la (H_f+P_f^2) R,\, R\ra \\
   & \geq - 8\alpha^3 \Re \eta_1\bar\eta_2
 \|\ceu\|_*^2
   - 8\alpha^3 \Re \eta_2\bar\eta_3 \|\cet\|_*^2 \\
   &
   - 8 \alpha^4 \Re \tilde{\eta_2}\bar\eta_1 \la \cedd,\,
     \ceddu \ra_*
   - 8 \alpha^4\Re\tilde\eta_2 \bar\eta_3\la
       \cedd,\, \ceddd\ra_*
   - 8 \alpha^4 \Re\eta_3\bar\eta_4 \la\cequ, \ceq\ra_* \\
   &
   - 8 \alpha^2 \Re \eta_1 \la \ceddu,\, R_2\ra_*
   - 8 \alpha^2 \Re \eta_3 \la \ceddd,\, R_2\ra_*
   - 8 \Re \eta_3 \alpha^2 \la \cequ,\,  R_4\ra_*\\
   &
   - c(1 + |\eta_2|^2 + |\tilde\eta_2|^2 + |\eta_4|^2)\alpha^5\, .
\end{split}
\end{equation}
\end{lemma}
%%%%%%%%%%%%%%%%%%%%%%%%%%%%%%%%%%%%%%%%%%%%%%%%
\textit{Proof.} Using the decomposition
\eqref{eq:decomp-1}-\eqref{eq:cond-1} of the ground state $\gs$,
we obtain
\begin{equation*}
\begin{split}
   & \Re\!\la \gs ,\, 4 \alpha^\frac12 P_f\!\cdot\!\Aa\gs\ra
 = \\
   & \Re\!\la R_1,\, (4\alpha^\frac12  P_f\!\cdot\!
 \Aa\eta_2\alpha\ced
 + 4\alpha^\frac12  P_f\!\cdot\!\Aa
  \tilde\eta_2\alpha^2\cedd
 + 4\alpha^\frac12 P_f\!\cdot\! \Aa R_2 )\, \ra \\
   & + \Re\!\la 2\eta_1\alpha^\frac32\ceu,
 \, (4\alpha^\frac12 P_f\!\cdot\! \Aa
  \eta_2 \alpha\ced
  + 4\alpha^\frac12 P_f\!\cdot\!
  \Aa\tilde\eta_2 \alpha^2 \cedd
  + 4\alpha^\frac12 P_f\!\cdot\!\Aa
  R_2)\, \ra \\
   & +
  \Re\!\la\eta_2\alpha^2\ced,\,
  ( 4 \alpha^\frac12  P_f\!\cdot\! \Aa 2
  \eta_3 \alpha^\frac32 \cet
  +  4 P_f\!\cdot\! \Aa R_3 )\, \ra \\
    &
  +\Re\!\la\tilde\eta_2\alpha^2\cedd,\,
  ( 4\alpha^\frac12 P_f\!\cdot\!\Aa
  2\eta_3\alpha^\frac32\cet
  +  4\alpha^\frac12 P_f\!\cdot\!\Aa R_3 )\, \ra \\
   & + \Re\!\la R_2,\,
 ( 4\alpha^\frac12 P_f\!\cdot\!\Aa 2\eta_3
  \alpha^\frac32\cet
 + 4\alpha^\frac12 P_f\!\cdot\!\Aa R_3 )\, \ra \\
   &  +  \Re\!\la 2\eta_3\alpha^\frac32\cet,\,
  ( 4\alpha^\frac12 P_f\!\cdot\! \Aa \eta_4 \alpha^2\ceq
  + 4\alpha^\frac12 P_f\!\cdot\!\Aa R_4 ) \, \ra \\
    & + \Re\! \la R_3,\,
 ( 4\alpha^\frac12 P_f\!\cdot\!\Aa \eta_4
  \alpha^2\ceq
  + 4\Re\alpha^\frac12 P_f\!\cdot\!\Aa R_4)\, \ra
  + \Re\!\la \Proj^{(\geq 4)}\gs,\, 4\alpha^\frac12 P_f\!\cdot\!\Aa
 \Proj^{(\geq 5)} \gs\ra \ .
\end{split}
\end{equation*}

For each value of $n$, we collect separately the terms in the
right hand side of this equality that stem from
$\Re\la\Proj^{(n)}\gs,\, 4 \alpha^\frac12
P_f\cdot\Aa\Proj^{(n+1)}\gs\ra$ . For estimating some of these
terms, like in \eqref{eq:argument} or \eqref{eq:argument2}, we
shall add a term like $\epsilon\la H_f R,\, R\ra$ or $\epsilon\la
P_f^2 R,\, R\ra$ borrowed from the left hand side of
\eqref{eq:main-appendix-1}.

- For $n=0$ there is no contribution.

- For $n=1$, we obtain the terms
\begin{equation}\label{eq:lem1-appendix-first}
\begin{split}
  \Re\la R_1,\, 4\alpha^\frac12  P_f\cdot \Aa\eta_2\alpha\ced\ra =
  -4 \alpha^\frac32 \Re \bar\eta_2\la R_1,\, \ceu\ra_* = 0\ ,
\end{split}
\end{equation}
where we used the $\la\cdot,\cdot\ra_*$-orthogonality of $R_1$ and
$\ceu$ given by \eqref{eq:cond-1},
\begin{equation}\label{eq:argument}
\begin{split}
    &
  \Re\la R_1,\, 4\alpha^\frac12  P_f\cdot\Aa
  \tilde\eta_2\alpha^2\cedd\ra + \frac{1}{16} \la P_f^2\,
  R_1,\, R_1\ra \\
    &
  \geq - 4 \| P_f  R_1\|\, \|\Aa\cedd\|\,
  \alpha^\frac52 |\tilde\eta_2| + \frac{1}{16}
  \| P_f R_1\|^2 \\
    & = (\frac14 \|P_f R_1\| - 8 \alpha^\frac52
  |\tilde\eta_2|\,  \|\Aa\cedd\|)^2 - 64 \|\Aa\cedd\|^2 \,
  |\tilde\eta_2|^2 \, \alpha^5\ \\
    & \geq - c |\tilde\eta_2|^2 \alpha^5\, .
\end{split}
\end{equation}
Note that we shall use the above argument several times in this
proof, as well as in the proof of the other lemmata of this
Appendix. We shall not give details again in these other cases.

We also have the following terms
\begin{equation}\label{eq:argument2}
\begin{split}
   & \Re\la R_1,\, 4\alpha^\frac12
 P_f\cdot \Aa R_2 \ra +\frac{1}{16}
 \la P_f^2\, R_1,\, R_1\ra + \frac{1}{16}\la H_f R_2,\, R_2\ra \\
   & \geq - c\alpha^\frac12 \|P_fR_1\|^2
  - c\alpha^\frac12 \|\Aa R_2\|^2
  +\frac{1}{16}
 \la P_f^2\, R_1,\, R_1\ra + \frac{1}{16}\la H_f R_2,\, R_2\ra\geq 0\
 ,
\end{split}
\end{equation}
where we used from \cite[Lemma~A4]{GLL} the inequality $\|\Aa
R_2\| \leq c \| H_f^\frac12 R_2\|$,
\begin{equation}
\begin{split}
    \Re\la 2\eta_1\alpha^\frac32\ceu,\, 4\alpha^\frac12 P_f\cdot \Aa
  \eta_2 \alpha\ced\ra & = 8\alpha^3 \Re\eta_1 \bar\eta_2\la\ceu,\,
  P_f\cdot\Aa \ced\ra \\
    & = -8\alpha^3 \Re\eta_1\bar\eta_2 \|\ceu\|_*^2\ ,
\end{split}
\end{equation}
\begin{equation}
\begin{split}
    \Re\la 2\eta_1\alpha^\frac32\ceu,\, 4\alpha^\frac12 P_f\cdot
  \Aa\tilde\eta_2 \alpha^2 \cedd\ra & = 8\alpha^4 \Re\eta_1
  \bar{\tilde{\eta_2}} \la \ceu,\, P_f\cdot \Aa\cedd\ra \\
    & = - 8 \alpha^4 \Re \eta_1 \bar{\tilde{\eta_2}} \la \ceddu,\,
  \cedd\ra_*\ ,
\end{split}
\end{equation}
and
\begin{equation}
\begin{split}
 \Re\la 2\eta_1 \alpha^\frac32\ceu,\,
  4\alpha^\frac12 P_f\cdot\Aa
  R_2\ra & = 8\alpha^2 \Re\eta_1 \la (H_f+P_f^2) P_f\cdot \Ac
  \ceu,\, \pr^\perp R_2\ra_* \\
    & = - 8\alpha^2 \Re\eta_1\la\ceddu,\, R_2\ra_*\, .
\end{split}
\end{equation}

- For $n=2$, we obtain the terms
\begin{equation}
\begin{split}
  \Re\la \eta_2 \alpha \ced, \, 4 \alpha^\frac12  P_f\cdot \Aa 2
  \eta_3 \alpha^\frac32 \cet\ra = -8\alpha^3 \Re\eta_2\bar\eta_3
  \|\cet\|_*^2\, ,
\end{split}
\end{equation}
\begin{equation}
\begin{split}
  \Re\la\eta_2\alpha^2\ced,\, 4 P_f\cdot \Aa R_3\ra =
  -4\alpha^\frac32 \Re\eta_2 \la\cet,\, R_3\ra_*=0\, ,
 \end{split}
\end{equation}
\begin{equation}
\begin{split}
  \Re\la\tilde\eta_2\alpha^2\cedd,\, 4\alpha^\frac12 P_f\cdot\Aa
  2\eta_3\alpha^\frac32\cet\ra = - 8\alpha^4\Re\tilde\eta_2\bar\eta_3
  \la\cedd,\, \ceddd\ra_* \, ,
\end{split}
\end{equation}
\begin{equation}
\begin{split}
  \Re\la \eta_2 \alpha^2 \cedd,\, 4\alpha^\frac12 P_f\cdot\Aa R_3\ra
  + \frac{1}{32} \la P_f^2 R_3,\, R_3\ra \geq
  - c|\eta_2|^2 \alpha^5\,
  ,
\end{split}
\end{equation}
where we used from \cite[Lemma~A4]{GLL} that $\| \Aa R_3 \| \leq c
\|H_f^\frac12 R_3 \|$,
\begin{equation}
\begin{split}
  \Re \la R_2,\, 4\alpha^\frac12 P_f\cdot\Aa 2\eta_3
  \alpha^\frac32\cet\ra
  = - 8\alpha^2 \Re \bar\eta_3\la R_2,\, \ceddd\ra_*\, ,
\end{split}
\end{equation}
and
\begin{equation}
\begin{split}
  \Re\la R_2,\, 4\alpha^\frac12 P_f\cdot\Aa R_3\ra +
  \frac{1}{16} \la P_f^2\, R_2,\, R_2\ra + \frac{1}{32}\la H_f R_3,\,
  R_3\ra \geq 0\ ,
\end{split}
\end{equation}
with similar argument as for\eqref{eq:argument2} for the last
inequality.

- For $n=3$, we obtain the terms

\begin{equation}
\begin{split}
  \Re\la 2\eta_3\alpha^\frac32\cet,\, 4\alpha^\frac12 P_f\cdot \Aa
  \eta_4 \alpha^2\ceq\ra = - 8\alpha^4\Re \eta_3\bar\eta_4 \la
  \cequ,\, \ceq\ra_*\, ,
\end{split}
\end{equation}
\begin{equation}
\begin{split}
  \Re\la2\eta_3\alpha^\frac32\cet,\, 4\alpha^\frac12 P_f\cdot\Aa
  R_4\ra
  = - 8\Re \eta_3 \alpha^2 \la\cequ,\, R_4\ra_*\, ,
\end{split}
\end{equation}
\begin{equation}
\begin{split}
  \Re \la R_3,\, 4\alpha^\frac12 P_f\cdot\Aa \eta_4
  \alpha^2\ceq\ra
  + \frac{1}{32} \la P_f^2\, R_3,\, R_3\ra \geq - c|\eta_4|^2
  \alpha^5\, ,
\end{split}
\end{equation}
and
\begin{equation}
\begin{split}
  \Re\la R_3,\, 4\Re\alpha^\frac12 P_f\cdot\Aa R_4\ra
  +\frac{1}{32} \la P_f^2 R_3,\, R_3\ra + \frac{1}{32}\la
  H_fR_4,\,R_4\ra \geq 0\, .
\end{split}
\end{equation}
- All contributions to the terms with $n\geq4$, give
\begin{equation}\label{eq:appendix-lem-1-last}
\begin{split}
 & \Re\la \Proj^{(\geq 4)}\gs,\, 4\alpha^\frac12 P_f\cdot\Aa
 \Proj^{(\geq 5)} \gs\ra + \frac{1}{16} \la H_f \Proj^{(\geq 5)}\gs,
 \Proj^{(\geq 5)}\gs\ra \\
 & \geq - c \alpha \|P_f \Proj^{(\geq
 4)}\gs\|^2 =- c \alpha^5 (1 + |\tilde{\eta_2}|^2 + |\eta_4|^2)\, ,
\end{split}
\end{equation}
where we used \eqref{eq:field-energy-bound} of
Proposition~\ref{prop:field-energy-bound}.

Collecting the inequalities
\eqref{eq:lem1-appendix-first}-\eqref{eq:appendix-lem-1-last}
conclude the proof of the Lemma. %\qed
%%%%%%%%%%%%%%%%%%%%%%%%%%%%%%%%%%%%%%%%%%%%%%%%

%%%%%%%%%%%%%%%%%%%%%%%%%%%%%%%%%%%%%%%%%%%%%%%%

%%%%%%%%%%%%%%%%%%%%%%%%%%%%%%%%%%%%%%%%%%%%%%%%
\subsection{Estimate of the term $\Re\la\gs,\, 2 \alpha\Aa\cdot\Aa\gs\ra$}
%%%%%%%%%%%%%%%%%%%%%%%%%%%%%%%%%%%%%%%%%%%%%%%%
\begin{lemma}\label{lem:appendix-2}
We have
\begin{equation}\label{eq:main-appendix-2}
\begin{split}
   &
  \Re\la\gs,\,2 \alpha \Aa\cdot\Aa\gs\ra
  + \frac18 \la (H_f +P_f^2) R,\, R\ra \\
  &
  \geq
  - 2\,\alpha^2 \Re \bar\eta_2 \|\ced\|_*^2
  - 8\, \alpha^2 \Re\eta_2 \la \ceqd,\, R_4\ra_*\\
  &
  + 8\, \alpha^4 \Re \eta_1\bar\eta_3 \la \ceu,\, \Aa\cdot\Aa\cet\ra
  - 8\, \alpha^4 \Re \eta_2\bar\eta_4 \la \ceqd,\, \ceq\ra_*\\
  &
  - c (1 + |\eta_1|^2 + |\eta_3|^2 + |\tilde\eta_2|^2
  + |\eta_4|^2 )\alpha^5
\end{split}
\end{equation}
\end{lemma}
%%%%%%%%%%%%%%%%%%%%%%%%%%%%%%%%%%%%%%%%%%%%%%%%
\textit{Proof.} Using the decomposition
\eqref{eq:decomp-1}-\eqref{eq:cond-1} of the ground state $\gs$
yields
\begin{equation*}
\begin{split}
   & \Re \la \gs,\, 2\alpha\Aa\cdot\Aa\gs\ra\\
   & =
 \Re\la \vac,\, ( 2\alpha\Aa\cdot\Aa\eta_2\alpha\ced
 + 2 \alpha\Aa\cdot\Aa \tilde\eta_2 \alpha^2\cedd
 +  2\alpha\Aa\cdot\Aa R_2 ) \, \ra \\
   & +
 \Re\la 2\eta_1\alpha^\frac32\ceu,\,
 ( 2\alpha\Aa\cdot\Aa
 2\eta_3\alpha^\frac32\cet
 +
 2 \alpha\Aa\cdot\Aa R_3 ) \, \ra \\
   & +
 \Re\la R_1,\,
 ( 2\alpha \Aa\cdot\Aa 2\eta_3\alpha^\frac32\cet
 +  2\alpha\Aa\cdot\Aa R_3 )\, \ra
 \\
   & +
 \Re\la \eta_2\alpha\ced,\,
 ( 2\alpha\Aa\cdot\Aa\eta_4\alpha^2\ceq
 + 2\alpha\Aa\cdot\Aa R_4 ) \, \ra \\
   & +
 \Re \la \tilde\eta_2\alpha^2 \cedd,\,
 ( 2\alpha\Aa\cdot\Aa
 \eta_4\alpha^2 \ceq
 + 2\Re\alpha\Aa\cdot\Aa R_4)\, \ra \\
   & +
 \Re\la R_2,\,
  ( 2\alpha \Aa\cdot\Aa\eta_4\alpha^2\ceq
  + 2\alpha\Aa\cdot\Aa R_4 ) \, \ra
 + \Re \la 2\eta_3\alpha^\frac32 \cet,\,
 2\alpha\Aa\cdot\Aa R_5 \ra \\
   &+ \Re \la R_3,\,  2\alpha\Aa\cdot\Aa R_5 ) \, \ra
 + \Re \la \Proj^{n\geq4}\gs,\, 2\alpha\Aa\cdot\Aa R_{\geq 6}\ra\,
 .
\end{split}
\end{equation*}
We collect in this expression the different contributions in
$\Re\la\Proj^{(n)}\gs,\, 2\alpha\Aa\cdot\Aa\Proj^{(n+2)}\gs\ra$
for each value of $n$. We shall use throughout this proof very
similar arguments to those used in the proof of
Lemma~\ref{lem:appendix-1}.

- For $n=0$, we have the terms
\begin{equation}\label{eq:lem2-appendix-first}
\begin{split}
 \Re\la \vac,\, 2\alpha\Aa\cdot\Aa\eta_2\alpha\ced\ra =
 - 2\, \alpha^2 \Re\eta_2 \|\ced\|_*^2\, ,
\end{split}
\end{equation}
\begin{equation}
\begin{split}
 \Re\la\vac,\, 2 \alpha\Aa\cdot\Aa \tilde\eta_2 \alpha^2\cedd\ra =
 -2\alpha^3 \Re\bar{\tilde{\eta_2}}\la\ced,\, \cedd\ra_* =0\, ,
\end{split}
\end{equation}
and
\begin{equation}
\begin{split}
 \Re\la\vac,\, 2\alpha\Aa\cdot\Aa R_2\ra = - 2\alpha\Re\la\ced,\,
 R_2\ra_*=0\, ,
\end{split}
\end{equation}

- For $n=1$, we have the terms
\begin{equation}
\begin{split}
 \Re\la 2\eta_1\alpha^\frac32\ceu,\, 2\alpha\Aa\cdot\Aa
 2\eta_3\alpha^\frac32\cet\ra
 = 8\, \alpha^4\Re \eta_1 \bar\eta_3 \la \ceu,\, \Aa\cdot\Aa
 \cet\ra\, ,
\end{split}
\end{equation}
\begin{equation}
\begin{split}
 \Re\la 2\eta_1\alpha^\frac32\ceu,2 \alpha\Aa\cdot\Aa R_3\ra +
 \frac{1}{32} \la H_f R_3,\, R_3\ra
 \geq - c |\eta_1|^2 \alpha^5\, ,
\end{split}
\end{equation}
\begin{equation}
\begin{split}
   &
 \Re\la R_1,\, 2\alpha \Aa\cdot\Aa 2\eta_3\alpha^\frac32\cet\ra
 + \frac{1}{32} \la H_f R_1,\, R_1\ra\\
   &
 = \Re\la H_f^\frac12 R_1,\, H_f^{-\frac12} 2 \alpha \Aa\cdot\Aa
 2\eta_3\alpha^\frac32\cet\ra + \frac{1}{32} \la H_f R_1,\, R_1\ra
 \geq - c|\eta_3|^2\alpha^5\, ,
\end{split}
\end{equation}
and, using \eqref{eq:estimate-N-theta} of
Proposition~\ref{prop:bound-2} and \cite[Lemma~A4]{GLL}
\begin{equation}
\begin{split}
   & \Re\la R_1, 2\alpha\Aa\cdot\Aa R_3\ra + \frac{1}{32}\la H_f
 R_3,\, R_3\ra^5
 \geq
 - c\alpha \|\Ac R_1\|\, \|H_f^\frac12 R_3\|
 +\frac{1}{32} \|H_f^\frac12 R_3\|^2 \\
 &
 \geq (-c\alpha^2\|\Ac R_1\|^2 -\frac{1}{32}\|H_f^\frac12 R_3\|^2) +
 \frac{1}{32} \| H_f^\frac12 R_3 \|^2 \\
   & \geq - c \alpha^2 (\|R_1\|^2 + \|H_f^\frac12 R_1\|^2) \geq -
 c\alpha^5 (1 + |\tilde{\eta_2}|^2 + |\eta_4|^2)
\end{split}
\end{equation}

- For $n=2$, we have the terms
\begin{equation}
\begin{split}
 \Re\la \eta_2\alpha\ced,\,
 2\alpha\Aa\cdot\Aa\eta_4\alpha^2\ceq\ra
 = - 8 \alpha^4  \Re\eta_2\bar\eta_4 \la\ceqd,\, \ceq\ra_*\, ,
\end{split}
\end{equation}
\begin{equation}
\begin{split}
 \Re\la\eta_2 \alpha\ced,\, 2\alpha\Aa\cdot\Aa R_4\ra =
 - 8 \alpha^2 \Re\eta_2\la\ceqd,\, R_4\ra_*\, ,
\end{split}
\end{equation}
\begin{equation}
\begin{split}
 \Re \la \tilde\eta_2\alpha^2 \cedd,\, 2\alpha\Aa\cdot\Aa
 \eta_4\alpha^2 \ceq\ra \geq - c\,\alpha^5
 (|\tilde\eta_2|^2+|\eta_4|^2)
\end{split}
\end{equation}
\begin{equation}
\begin{split}
 \Re\la\tilde\eta_2\alpha^2\cedd,\, 2\Re\alpha\Aa\cdot\Aa R_4\ra
 \geq - c\, \alpha^5\, (1+ |\tilde\eta_2|^2 + |\eta_4|^2) ,\,
\end{split}
\end{equation}
using $\|\Aa R_4\|\leq c\,\|H_f^\frac12 R_4\|  \leq c \alpha^5 (1+
|\tilde{\eta_2}|^2 + |\eta_4|^2)$ (respectively
\cite[Lemma~A4]{GLL} and
Proposition~\ref{prop:field-energy-bound}). We also have the terms
\begin{equation}
\begin{split}
   &
 \Re\la R_2,\, 2\alpha \Aa\cdot\Aa\eta_4\alpha^2\ceq\ra +
 \frac{1}{32}\la H_fR_2,\, R_2\ra \\
   &
 =
 \Re\la H_f^\frac12R_2,\, 2\alpha H_f^{-\frac12}
 \Aa\cdot\Aa\eta_4\alpha^2\ceq\ra +
 \frac{1}{32}\la H_fR_2,\, R_2\ra \geq - c |\eta_4|^2\alpha^6
\end{split}
\end{equation}
and
\begin{equation}\label{eq:56}
\begin{split}
   &
 \Re\la R_2, 2\alpha\Aa\cdot\Aa R_4\ra + \frac{1}{32} \la
 H_fR_4,\, R_4\ra
 \geq -c \alpha^2 \|\Ac R_2\|^2 \\
   &
 \geq\! - c \alpha^2 (\|H_f R_2\|^2 \!+\! \|R_2\|^2)
 \geq \! - c \alpha^2 (\|R_2\|_*^2\! +\! \| \Theta \|^2
 \! + \! |\tilde{\eta_2}|^2 \| \alpha^2 \cedd\|^2
 \! + \! |\eta_4|^2 \| \alpha^2 \ceq \|^2   ) \\
   & \geq - c \alpha^5 (1 + |\tilde{\eta_2}|^2 +
 |\eta_4|^2) ,
\end{split}
\end{equation}
for $\Theta$ defined by \eqref{eq:def-theta} and using from
\eqref{eq:bound-2} of Corollary~\ref{corollary-1} that
$\|\Theta\|^2 = \Or(\alpha^3)$ and from
\eqref{eq:field-energy-bound} of
Proposition~\ref{prop:field-energy-bound} that $\|R_2\|_*^2 \leq c
\alpha^4 (1 + |\tilde{\eta_2}|^2 + |\eta_4|^2)$.

- For $n\geq 3$ we collect all the terms as follows
\begin{equation}
\begin{split}
 \Re \la 2\eta_3\alpha^\frac32 \cet,\, 2\alpha\Aa\cdot\Aa R_5\ra +
 \frac{1}{32} \la H_f R_5,\, R_5\ra \geq - c|\eta_3|^2 \alpha^5\, ,
\end{split}
\end{equation}
\begin{equation}
\begin{split}
 \Re \la R_3,\, 2\alpha\Aa\cdot\Aa R_5\ra +
 \frac{1}{32} \la H_f R_5,\, R_5\ra \geq - c \alpha^5\, ,
\end{split}
\end{equation}
using \eqref{eq:bound-2} of Corollary~\ref{corollary-1} in the
last inequality. Finally, we get
\begin{equation}\label{eq:lem2-appendix-last}
\begin{split}
   \Re \la \Proj^{n\geq4}\gs,\, 2\alpha\Aa\cdot\Aa R_{\geq 6}\ra
 +
 \frac{1}{32} \la H_f R_{\geq 6},\, R_{\geq 6}\ra
 \geq - c \alpha^5 (1 + |\tilde{\eta_2}|^2 + |\eta_4|^2) \, ,
\end{split}
\end{equation}
Collecting
\eqref{eq:lem2-appendix-first}-\eqref{eq:lem2-appendix-last}
yields
the result. %\qed

%%%%%%%%%%%%%%%%%%%%%%%%%%%%%%%%%%%%%%%%%%%%%%%%
\subsection{Estimate of the term $\la\gs,\,2 \alpha \Ac\cdot\Aa\gs\ra$}

%%%%%%%%%%%%%%%%%%%%%%%%%%%%%%%%%%%%%%%%%%%%%%%%
\begin{lemma}\label{lem:appendix-3}
We have
\begin{equation}\label{eq:main-appendix-3}
\begin{split}
   & \la\gs,\,2 \alpha \Ac \! \cdot \! \Aa\gs\ra
 \! +\! \frac18 \la (H_f +P_f^2) R,\, R\ra
 \geq
 - 8 \alpha^2\Re \bar\eta_2 \la R_2, \ceddt\ra_*
 + 2 |\eta_2|^2\alpha^3 \|\Aa\ced\|^2\\
   &
 + 8 \alpha^4 |\eta_1|^2 \|\Aa \ceu\|^2
 + 8 \alpha^4 |\eta_3|^2 \|\Aa \cet\|^2
 - 8 \alpha^4 \Re \tilde\eta_2 \eta_2 \la \cedd,\, \ceddt\ra_*
 \\
   &
 - c\alpha^5( 1 + |\eta_1|^2 + |\eta_3|^2 + |\tilde\eta_2|^2)\, .
\end{split}
\end{equation}
\end{lemma}
%%%%%%%%%%%%%%%%%%%%%%%%%%%%%%%%%%%%%%%%%%%%%%%%
\textit{Proof.} With the decomposition
\eqref{eq:decomp-1}-\eqref{eq:cond-1} of $\gs$ we get
\begin{equation*}
 \begin{split}
   & \la \gs,\, 2\alpha\Ac\!\cdot\!\Ac\gs\ra
 =
 \la 2\eta_1 \alpha^\frac32 \ceu,\,
  2\alpha\Ac\!\cdot\!\Aa 2\eta_1\alpha^\frac32\ceu \ra
  +
  2 \Re \la 2\eta_1 \alpha^\frac32 \ceu,\,
  2\alpha\Ac\!\cdot\!\Aa R_1  \, \ra \\
    &
  + \la\eta_2\alpha\ced,\,
  2\alpha\Ac\!\cdot\!\Aa \eta_2\alpha\ced \ra
  + 2\Re \la\eta_2\alpha\ced,\,
  2\alpha\Ac\!\cdot\!\Aa \tilde\eta_2 \alpha^2 \cedd  \, \ra \\
    & + \la \tilde\eta_2\alpha^2\cedd,\,
  2\alpha\Ac\!\cdot\!\Aa\tilde\eta_2 \alpha^2 \cedd \ra
  + \la R_2,\, 2\alpha\Ac\!\cdot\!\Aa R_2\ra
  + 2\Re\la R_2,\, 2\alpha\Ac\!\cdot\!\Aa \eta_2\ced\ra \\
    &
  + 2\Re \la R_2, 2\alpha \Ac\!\cdot\!\Aa
  \tilde\eta_2\alpha^2\ceddd\ra
  + \la 2\eta_3\alpha^\frac32 \cet,\, 2\alpha\Ac\!\cdot\!\Aa
  2\eta_3\alpha^\frac32\cet\ra \\
    &
  + 2\Re\la 2\eta_3\alpha^\frac32\cet,
  2\alpha\Ac\!\cdot\!\Aa R_3\ra
  + \la R_3,\, 2\alpha\Ac\cdot\Aa R_3\ra
  + \la \Proj^{(\geq 4)}\gs,\,
  2\alpha\Ac\cdot\Aa\Proj^{(\geq4)}\gs\ra \, .
\end{split}
\end{equation*}

For each value of $n$, we next collect in the above equality the
different contributions of $\la\Proj^{(n)}\gs,\,
2\alpha\Ac\cdot\Aa\Proj^{(n)}\gs\ra$.

- For $n=0$, there is no term.

- For $n=1$, we have
\begin{equation}\label{eq:lem3-appendix-first}
\begin{split}
 \la 2\eta_1 \alpha^\frac32 \ceu,\, 2\alpha\Ac\cdot\Aa
 2\eta_1\alpha^\frac32\ceu\ra
 = 8 |\eta_1|^2 \alpha^4 \|\Aa\ceu\|^2\, ,
\end{split}
\end{equation}
and
\begin{equation}
\begin{split}
 2\Re \la 2\eta_1\alpha^\frac32\ceu,\, 2\alpha\Ac\cdot\Aa R_1\ra
 +\frac{1}{32} \la H_f R_1,\, R_1\ra
 \geq - c |\eta_1|^2 \alpha^5\, .
\end{split}
\end{equation}

- For $n=2$, we obtain
\begin{equation}
\begin{split}
 \la\eta_2\alpha\ced,\, 2\alpha\Ac\cdot\Aa \eta_2\alpha\ced\ra
 = 2 |\eta_2|^2 \alpha^3 \|\Aa\ced\|^2\, ,
\end{split}
\end{equation}
\begin{equation}
\begin{split}
 2 \Re\la\eta_2 \alpha\ced,\, 2\alpha\Ac\cdot\Aa \tilde\eta_2
 \alpha^2 \cedd\ra
 = - 8 \Re \eta_2 \bar{\tilde{\eta_2}} \alpha^4 \la \ceddt,\,
 \cedd\ra_*\, ,
\end{split}
\end{equation}
\begin{equation}
\begin{split}
 \la \tilde\eta_2\alpha^2\cedd,\, 2\alpha\Ac\cdot\Aa\tilde\eta_2
 \alpha^2 \cedd\ra
 = 4 \alpha^5 |\tilde\eta_2|^2 \|\Aa\cedd\|^2\, ,
\end{split}
\end{equation}
\begin{equation}
\begin{split}
 \la R_2,\, 2\alpha\Ac\cdot\Aa R_2\ra \geq 0\, ,
\end{split}
\end{equation}
\begin{equation}
\begin{split}
 2\Re\la R_2,\, 2\alpha\Ac\cdot\Aa \eta_2\ced\ra
  = - 8\alpha^2\Re\bar\eta_2\la R_2,\, \ceddt\ra_*\, ,
\end{split}
\end{equation}
and
\begin{equation}
\begin{split}
 2\Re \la R_2, 2\alpha \Ac\cdot\Aa \tilde\eta_2\alpha^2\ceddd\ra
 + \frac{1}{32} \la H_f R_2,\, R_2\ra \geq -
 c|\tilde\eta_2|^2\alpha^5\, .
\end{split}
\end{equation}

- For $n=3$, we have
\begin{equation}
\begin{split}
 \la 2\eta_3\alpha^\frac32 \cet,\, 2\alpha\Ac\cdot\Aa
 2\eta_3\alpha^\frac32\cet\ra
 = 8 \alpha^4 |\eta_3|^2 \|\Aa\cet\|^2\, ,
\end{split}
\end{equation}
\begin{equation}
\begin{split}
 2\Re\la 2\eta_3\alpha^\frac32\cet, 2\alpha\Ac\cdot\Aa R_3\ra
 + \frac{1}{32} \la H_f R_3,\, R_3\ra \geq - |\eta_3|^2 \alpha^5\,
 ,
\end{split}
\end{equation}
and
\begin{equation}
\begin{split}
 \la R_3,\, 2\alpha\Ac\cdot\Aa R_3\ra \geq 0\, ,
\end{split}
\end{equation}

- For $n\geq 4$, we obtain
\begin{equation}\label{eq:lem3-appendix-last}
\begin{split}
 \la \Proj^{(\geq 4)}\gs,\,
 2\alpha\Ac\cdot\Aa\Proj^{(\geq4)}\gs\ra \geq 0\, .
\end{split}
\end{equation}

Collecting
\eqref{eq:lem3-appendix-first}-\eqref{eq:lem3-appendix-last}
concludes the proof of the lemma. %\qed

%%%%%%%%%%%%%%%%%%%%%%%%%%%%%%%%%%%%%%%%%%%%%%%%
\subsection{Computation of the term $\la\gs,\, H_f + P_f^2\gs\ra$}

%%%%%%%%%%%%%%%%%%%%%%%%%%%%%%%%%%%%%%%%%%%%%%%%
\begin{lemma}\label{lem:appendix-4}
We have
\begin{equation}\label{eq:main-appendix-4}
\begin{split}
   \la (H_f + P_f^2) \gs,\, \gs\ra
  = &
  \ |\eta_2|^2 \alpha^2 \|\ced\|_*^2
  + 4 |\eta_1|^2 \alpha^3 \|\ceu\|_*^2
  + 4 |\eta_3|^2 \alpha^3 \|\cet\|_*^2 \\
    & + |\tilde\eta_2|^2 \alpha^4\|\cedd\|_*^2
  + |\eta_4|^2\alpha^4 \|\ceq\|_*^2
  + \|R\|_*^2\, .
\end{split}
\end{equation}
\end{lemma}
\textit{Proof.} Using the decomposition \eqref{eq:decomp-1} of
$\gs$ and using the whole set of orthogonalities with respect to
$\la\cdot,\cdot\ra_*$ given in \eqref{eq:cond-1}, we obtain that
all crossed terms are zero. The proof is thus straightforward.
%\qed
%%%%%%%%%%%%%%%%%%%%%%%%%%%%%%%%%%%%%%%%%%%%%%%%

%\subsection{}
%\end{appendix}
%%%%%%%%%%%%%%%%%%%%%%%%%%%%%%%%%%%%%%%%%%%%%%%%

%%%%%%%%%%%%%%%%%%%%%%%%%%%%%%%%%%%%%%%%%%%%%%%%%%%%%%%%%%%%%
%%%%%%%%%%%%%%%%%%%%%%%%%%%%%%%%%%%%%%%%%%%%%%%%%%%%%%%%%%%%%
%%%%%%%%%%    BIBLIOGRAPHY   %%%%%%%%%%%%%%%%%%%%%%%%%%%%%%%%
%%%%%%%%%%%%%%%%%%%%%%%%%%%%%%%%%%%%%%%%%%%%%%%%%%%%%%%%%%%%%
%%%%%%%%%%%%%%%%%%%%%%%%%%%%%%%%%%%%%%%%%%%%%%%%%%%%%%%%%%%%%

\bibliographystyle{plain}

\end{document}